\documentclass[sigconf]{acmart}

\usepackage{savesym}
\savesymbol{checkmark}
\savesymbol{ding}
\usepackage{pifont}
\usepackage{dingbat}
\usepackage{fontawesome5}
\usepackage{tabularx}

\AtBeginDocument{%
  }

\setcopyright{acmlicensed}
\copyrightyear{2025}
\acmYear{2025}
\acmDOI{XXXXXXX.XXXXXXX}
\acmConference[Preprint]{Make sure to enter the correct
  conference title from your rights confirmation email}{April 2025}{arXiv}
\acmISBN{978-1-4503-XXXX-X/2018/06}

\makeatletter 
\renewcommand{\paragraph}{%
  \@startsection{paragraph}{4}{\z@}
                {1.5ex \@plus1ex \@minus.2ex}
                {-1em}
                {\normalfont\normalsize\bfseries\noindent}
}
\makeatother

\PassOptionsToPackage{dvipsnames}{xcolor}
\usepackage{multirow}
\usepackage{enumitem}

\setlist[itemize]{leftmargin=8mm}
\setlist[enumerate]{leftmargin=8mm}

\definecolor{myGreen}{HTML}{2E8B57}  
\definecolor{myYellow}{HTML}{DAA520} 
\definecolor{myRed}{HTML}{B22222}    

\newcommand{\mycheckmark}{\textcolor{myGreen}{\checkmark}}
\newcommand{\mydash}{\textcolor{myYellow}{--}}
\newcommand{\mypartial}{\faSpinner}
\newcommand{\mycross}{\textcolor{myRed}{\ding{55}}}

\newcommand{\pref}[1]{\mycheckmark #1}
\newcommand{\nopref}{\mydash} 
\newcommand{\notpref}[1]{\mycross #1}

\begin{document}

\title{Prompt Orchestration Markup Language}

\author{Yuge Zhang, Nan Chen, Jiahang Xu, Yuqing Yang}
\email{semantipy@microsoft.com}
\affiliation{%
  \institution{Microsoft Research}
  \city{Shanghai}
  \country{China}
}

\renewcommand{\shortauthors}{POML Team}

\begin{abstract}
    Large Language Models (LLMs) require sophisticated prompting, yet current practices face challenges in structure, data integration, format sensitivity, and tooling.
    Existing methods lack comprehensive solutions for organizing complex prompts involving diverse data types (documents, tables, images) or managing presentation variations systematically.
    To address these gaps, we introduce POML (Prompt Orchestration Markup Language).
    POML employs component-based markup for logical structure (roles, tasks, examples), specialized tags for seamless data integration, and a CSS-like styling system to decouple content from presentation, reducing formatting sensitivity.
    It includes templating for dynamic prompts and a comprehensive developer toolkit (IDE support, SDKs) to improve version control and collaboration.
    We validate POML through two case studies demonstrating its impact on complex application integration (PomLink) and accuracy performance (TableQA), as well as a user study assessing its effectiveness in real-world development scenarios.
\end{abstract}


\begin{teaserfigure}
  \centering
  \begin{minipage}[b]{0.77\textwidth}
      \centering
      \includegraphics[width=\linewidth]{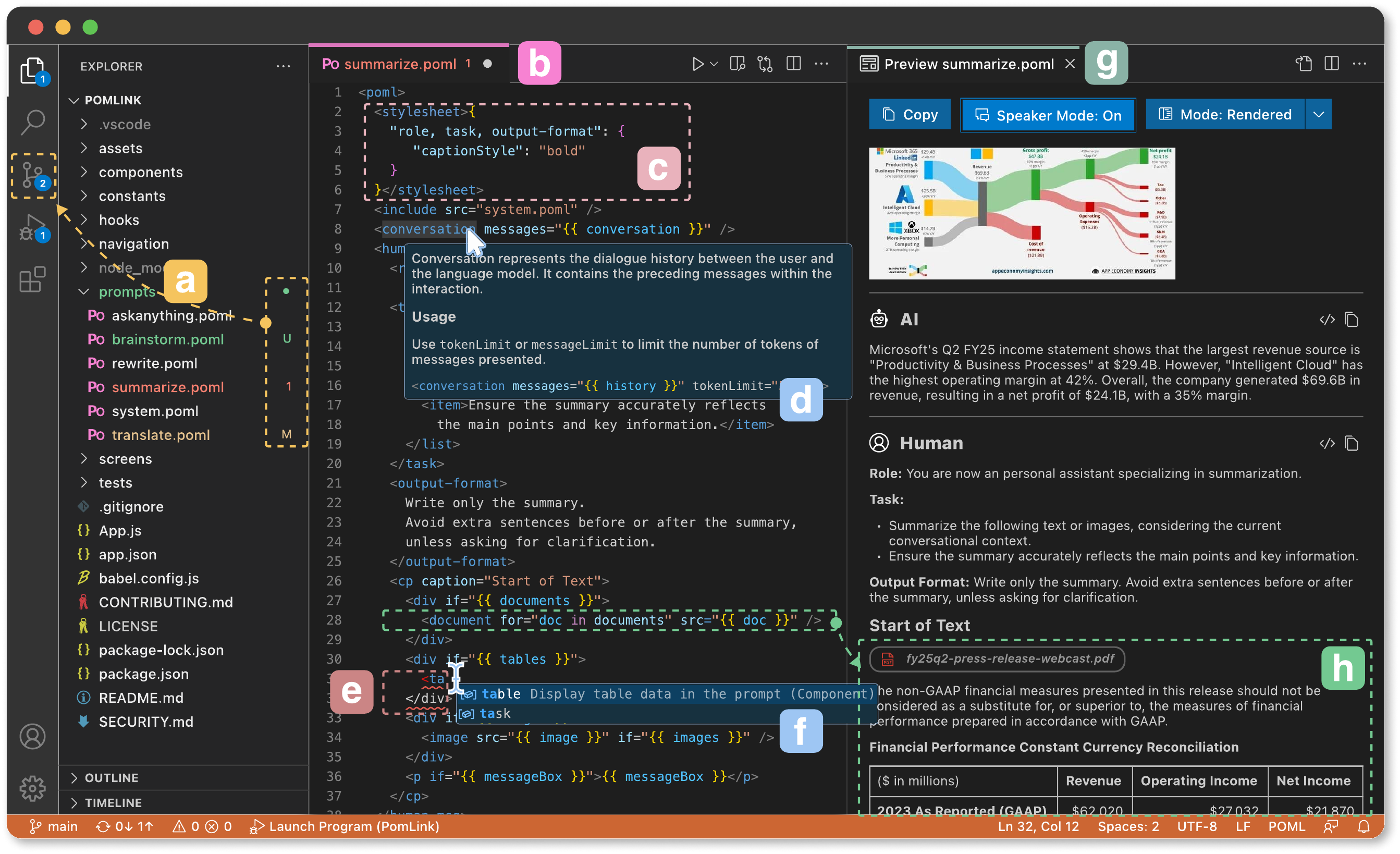}
      \vspace{-6mm}
      \caption*{(1)}
  \end{minipage}%
  \begin{minipage}[b]{0.224\textwidth}
      \centering
      \includegraphics[width=\linewidth]{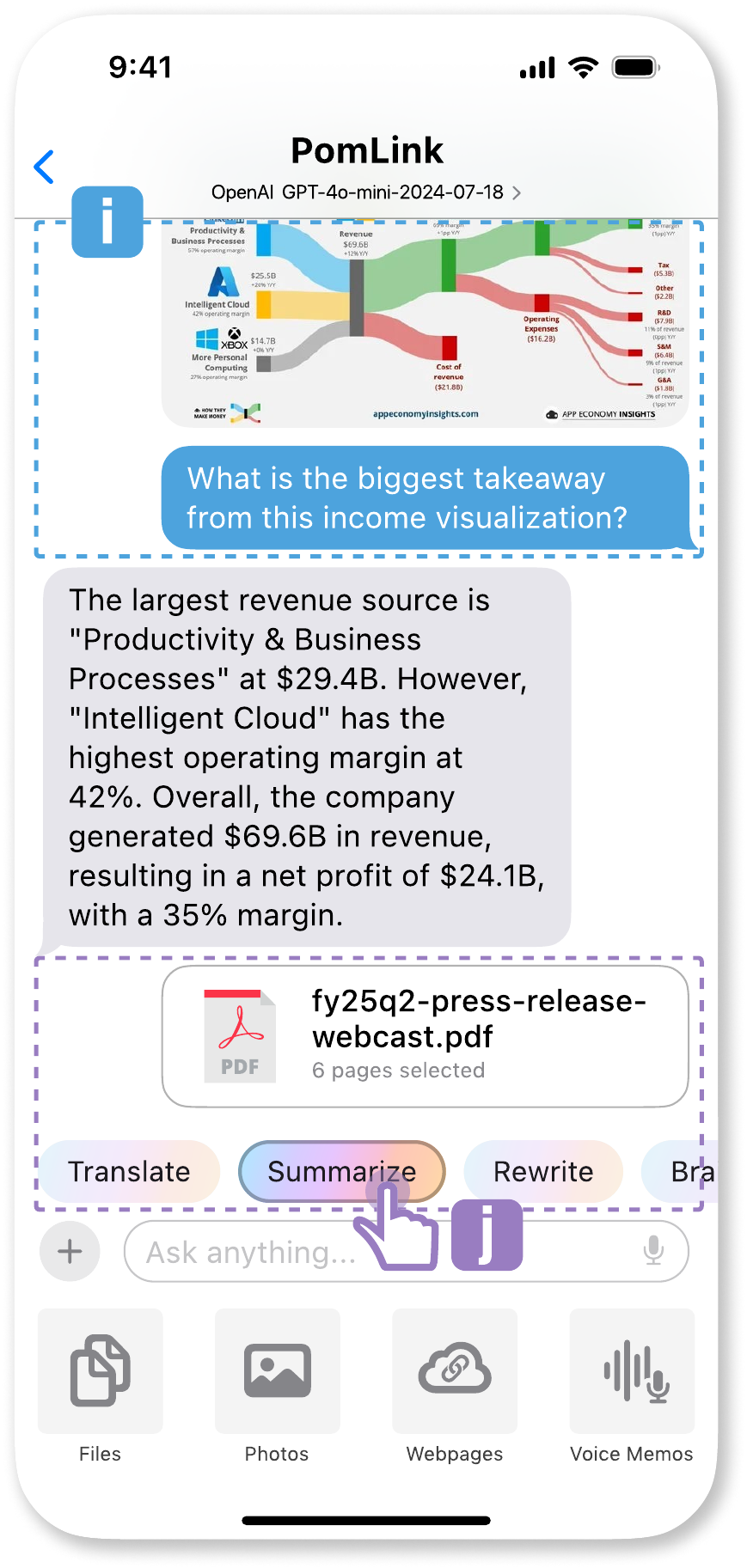}
      \vspace{-6mm}
      \caption*{(2)}
  \end{minipage}
  \vspace{-3mm}
  \caption{Developing the PomLink iOS prototype in VSCode -- a case study of POML.
  \textbf{(1)} The development environment in VSCode, showcasing structured prompt authoring with POML: \textbf{(a)} prompt version control with Git; \textbf{(b)} a POML file example; \textbf{(c)} stylesheet definition for presentation control; \textbf{(d)} hover documentation for inline assistance; \textbf{(e)} inline diagnostics for error checking; \textbf{(f)} context-aware auto-completion; \textbf{(g)} the live preview panel displaying; \textbf{(h)} rendered embedded data (e.g., documents and tables).
  \textbf{(2)} An application prototype developed with POML that structures multimodal inputs for LLM analysis, supporting tasks such as \textbf{(i)} querying visual content and \textbf{(j)} summarizing documents based on rich context.}
  \label{fig:teaser}
\end{teaserfigure}


\settopmatter{printacmref=false}
\maketitle

\section{Introduction}

Large Language Models (LLMs) have demonstrated remarkable capabilities across diverse tasks through carefully engineered prompts.
These models, exemplified by the GPT series \cite{brown2020language,openai2024gpt4technicalreport}, LLaMA \cite{grattafiori2024llama3herdmodels}, and others \cite{geminiteam2024gemini15unlockingmultimodal,deepseekai2024deepseekv3}, have shown proficiency in parsing and acting upon instructions in various domains.
Their applications span code generation \cite{chen2021evaluating}, question answering \cite{hendrycks2021measuring}, mathematical reasoning \cite{cobbe2021training}, and interactive agents \cite{shridar2021alfworld, yao2022webshop}.
As these applications grow in complexity, so too does the sophistication of prompting techniques used to elicit desired behaviors from LLMs.
Advanced prompting strategies such as chain-of-thought \cite{wei2022chain}, few-shot in-context learning \cite{brown2020language}, and ReAct (reasoning and acting) \cite{yaoreact} have emerged as critical methods for improving model performance.

Within this evolving landscape, ``structured prompting'' \cite{sahoo2024systematic, rtfprompt, wang2024langgpt} has gained prominence as a systematic approach to organizing instructions and context data within prompts.
These structured approaches offer numerous advantages: they componentize and standardize prompts for better automated editing \cite{schnabel2024prompts}, facilitate team collaboration \cite{promptml}, improve development efficiency \cite{chatxml}, support fine-grained editing \cite{desmond2024exploring}, enable reuse of proven designs \cite{wang2024langgpt}, and potentially boost performance \cite{li2024guiding}.
As LLM applications increasingly integrate multiple data types -- including images \cite{geminiteam2024gemini15unlockingmultimodal}, tables \cite{sui2024tablemeetsllmlarge}, and documents \cite{lin2024revolutionizingretrievalaugmentedgenerationenhanced}, the need for a more organized approach to prompt design becomes increasingly apparent.

Despite these advances, prompt engineering faces four significant challenges that impede effective development and deployment of LLM applications.
\emph{First,} current practices lack standardized methods for the \textbf{structured orchestration of prompts}.
Projects following ``structured prompting'' practices often involve scattered instructions, roles, tasks, and examples \cite{rtfprompt, wang2024langgpt}, hindering maintenance and collaboration, especially in complex projects.
\emph{Second,} \textbf{integrating diverse data sources} remains complex.
Large LLM applications increasingly require references to documents \cite{lin2024revolutionizingretrievalaugmentedgenerationenhanced}, tables \cite{zhang2024benchmarking, chen2024viseval}, or images \cite{liu2023visualinstructiontuning}.
Properly integrating and presenting these varied sources within prompts poses a significant engineering burden.
\emph{Third,} LLMs exhibit significant \textbf{sensitivity to formatting and presentation} \cite{sclarquantifying, salinas2024butterfly, voronov2024mind}.
Research has documented ``butterfly effects,'' where minor textual variations can dramatically alter results.
The absence of a decoupled styling layer complicates systematic testing and refinement of prompt variations \cite{liu2025promptcontentenhancingllm, schnabel2024symbolicpromptprogramsearch}.
\emph{Finally,} prompt development and management suffer from \textbf{inadequate tooling}.
Current workflows often lack effective version control and diff visibility \cite{desmond2024exploring}, making tracking changes and collaboration difficult.
Plain-text prompts are cumbersome to manage \cite{bach2022promptsource, zamfirescu2023johnny}, and limited IDE support hinders systematic improvement \cite{arawjo2024chainforge, strobelt2022interactive}.

Several existing approaches attempt to address these challenges, but each has its specific limitations.
Workflow and agent orchestration tools like LangChain \cite{langchain}, Microsoft Guidance \cite{microsoftguidance}, and PromptChainer \cite{wu2022promptchainer} manage multi-step flows but treat prompts as plain text or minimal templates, lacking structured data presentation or advanced styling capabilities.
Markup-based prompting approaches such as ChatML \cite{chatml}, MDXPrompt \cite{mdxprompt}, PromptML \cite{promptml}, and VSCode PromptTSX \cite{vscodeprompttsx} provide component-based frameworks but lack effective data handling, a well-separated styling system, or comprehensive development support.
User interfaces for prompt comparison or logging exist \cite{arawjo2024chainforge, mishra2023promptaid}, yet lack the fundamental tools to author or structure prompts before running comparisons or evaluations.

To address these multifaceted challenges comprehensively, we introduce \textbf{POML (Prompt Orchestration Markup Language)}.
POML provides a novel paradigm designed to bring structure, maintainability, and versatility to advanced prompt engineering.
POML is based on an HTML-like \textbf{structured markup language} featuring semantic components, such as \texttt{<role>}, \texttt{<task>}, and \texttt{<example>}, encouraging modular design and enhancing prompt reusability.
To ease data integration, POML incorporates specialized \textbf{data components} (e.g., \texttt{<document>}, \texttt{<table>}, \texttt{<img>}) that seamlessly embed or reference external data sources like text files, spreadsheets, and images \cite{chen2024viseval, xia2024agentless}.
These data components offer customizable formatting options and eliminate error-prone manual text merging.
To address format sensitivity, POML introduces a \textbf{styling system} inspired by CSS \cite{css, w3c}.
This decouples content from presentation, allowing developers to modify styling (e.g., verbosity, syntax format) via separate \texttt{<stylesheet>} definitions (\autoref{fig:teaser} (c)) and systematically experiment with format variations without altering core logic.
POML also includes a built-in \textbf{templating engine} for dynamically generating complex, data-driven prompts.

Beyond the language itself, POML addresses inadequate tooling with a comprehensive \textbf{development toolkit}.
This includes an Integrated Development Environment (IDE) extension (initially for Visual Studio Code, see \autoref{fig:teaser} (1)) providing essential development aids like real-time previews (g), inline diagnostics (e), and auto-completion (f).
The toolkit also provides Software Development Kits (SDKs) for seamless integration into Python and Node.js applications \cite{langchain, semantickernel}.
Overall, the toolkit streamlines the prompt engineering lifecycle, enhancing authoring efficiency, debugging, version control (\autoref{fig:teaser} (a)), collaboration, and the systematic evolution of prompts.

We validate the effectiveness and utility of POML through rigorous empirical evaluation.
This includes two distinct case studies and a formal user study.
The first case study showcases POML's practical application in building PomLink -- an iOS agent application prototype (\autoref{fig:teaser} (2)).
This study highlights how POML's data components, styling, and tooling enabled the rapid development (a functional prototype in two days) of a complex application prototype requiring integration of documents, images, and tables.
The second case study, focusing on Table Question Answering (TableQA), systematically explores the impact of prompt styling using POML's styling system.
It demonstrates that stylistic variations, easily managed by POML, can cause dramatic, model-specific differences in LLM performance (e.g., accuracy improvements over 9x for some models) on table-based reasoning tasks \cite{pasupat2015compositional}.
Finally, our user study involving participants with varying expertise assessed POML's usability across five representative tasks \cite{zamfirescu2023johnny, dang2022prompt}.
The results confirmed POML's effectiveness in structuring prompts and handling data, with participants particularly valuing the data components and development toolkits, while also providing constructive feedback for future enhancements.

The primary contributions of this work are:
\begin{enumerate}
    \item POML: A novel markup language designed specifically for prompt engineering, with specialized data components for effective integration of diverse data types, and a styling system for decoupling prompt content from presentation, enabling systematic control over formatting and mitigating LLM format sensitivity.
    \item An integrated suite of tools including an IDE extension with features like syntax highlighting, live preview, diagnostics, interactive testing, and multi-language SDKs to improve the development workflow, improve version control, and facilitate project management and collaboration.
    \item POML's practical benefits and impact are empirically validated through two detailed case studies and a formal user study that assessed usability and effectiveness in realistic scenarios.
\end{enumerate}

\section{Related Works}

\subsection{Prompt Engineering and Structured Prompt}

Large Language Models (LLMs) demonstrate remarkable capabilities, yet their performance is highly sensitive to the input prompt's quality and format \cite{sahoo2024systematic}.
This sensitivity is particularly pronounced when prompts involve complex data types such as tables, charts, or images, where carefully tuned presentation is often necessary for reliable results \cite{zhang2024benchmarking, chen2024viseval, xia2024agentless, fatemitalk, he2025tableloralowrankadaptationtable}.
Indeed, research highlights a ``butterfly effect,'' where minor variations in phrasing, formatting, markup, or content order can dramatically alter LLM outputs and accuracy \cite{salinas2024butterfly, voronov2024mind, he2024does, liu2025promptcontentenhancingllm, zhuo2024prosa}.

Consequently, prompt engineering -- the practice of designing inputs to optimize LLM outputs -- has become essential \cite{sahoo2024systematic}.
Research has shown that enhancing prompts with in-context examples or iterative reasoning steps significantly improves results \cite{brown2020language, wei2022chain, dhuliawala2024chain, wangself, yaoreact}.
Within this field, structured prompting has emerged as a key practice, organizing prompts into standardized roles, tasks, or program-like formats \cite{structuredprompt, rtfprompt, chatgpt3freepromptlist, wang2024langgpt, schnabel2024prompts}.
Such approaches offer advantages including improved reusability and maintainability \cite{schnabel2024prompts,promptml,chatxml,desmond2024exploring,wang2024langgpt,li2024guiding}.
Conversely, unstructured prompts often impede team-based workflows, making targeted modifications difficult without affecting other prompt sections \cite{desmond2024exploring}.
The lack of clear boundaries or standardized documentation for prompt segments (e.g., roles, tasks, examples) hinders the development of shared tooling for parsing, manipulating, or reusing prompt components programmatically.

\subsection{Prompt Development Challenges and Tooling}
\label{sec:related_challenges_tooling}

\paragraph{Studies on Prompt Usage and Challenges}
Studies on prompt usage patterns reveal significant challenges faced by users, especially non-experts.
Minimalist web/chat UIs like OpenAI's Playground and ChatGPT \cite{openaiplayground, openaichatgpt} serve as short-term sandboxes, leaving a gap between experimental prototyping and large-scale and production-ready prompt designs.
In industry, trial-and-error approaches are common but hinder systematic improvements, as users often make erroneous assumptions about how to communicate with LLMs \cite{zamfirescu2023johnny, dang2022prompt}.
This continuous editing process also makes prompt history tracking difficult, complicating version control and collaboration.
Text-based prompts frequently become opaque or difficult to manage over time \cite{bach2022promptsource, desmond2024exploring}.
Tagging approaches are being explored to alleviate these issues by improving differential comparisons and collaborative editing \cite{wang2024langgpt, promptml}, aligning with structured prompting practices.

\paragraph{Prompt Toolkits for Development and Evaluation}
Several integrated prompt toolkits have been developed to assist users in prompt engineering.
Flow-based or multi-prompt systems provide pipelines for chaining multiple prompts with optional retrieval components in complex LLM applications \cite{langchain, wu2022promptchainer, microsoftguidance, semantickernel}.
Other tools focus on prompt management and version control through standalone or web-based services \cite{latitude-llm, pezzo}.
However, these systems typically manage the high-level orchestration of multiple steps or prompts, often treating individual prompts as opaque strings or simple templates, rather than providing detailed frameworks for structuring the content \emph{within} a single complex prompt.
POML can integrate with these systems as a specialized syntax for enhanced clarity and structure within individual prompt steps.
Another stream of existing works is prompt evaluation interfaces, offering UIs for comparing or perturbing prompts \cite{arawjo2024chainforge, mishra2023promptaid, strobelt2022interactive, promptfoo, langsmith}.
These are largely \emph{orthogonal} to POML's focus on prompt authoring and structure, and could potentially use POML's declared sub-blocks for more fine-grained testing and analysis.

\subsection{Web Frameworks and Prompt Markup Languages}
\label{sec:related_markup_component}

\paragraph{Inspiration from Web Frameworks and Programming}
The evolution of web development offers relevant parallels and insights for prompt engineering.
Both domains share concerns around clarity, dynamic content, and adaptable layouts -- whether for different screen sizes or context windows \cite{promptdesign}.
Principles like the separation of structure (HTML \cite{html, html1991}), presentation (CSS \cite{css, w3c}), and behavior (JavaScript \cite{javascript}) have proven crucial for managing complexity in web development, providing a blueprint for prompt frameworks.
Early dynamic websites relied on templating engines to map data into presentation views \cite{minamide2005static, django, jinja, twig, handlebars}, akin to simple prompt templating in tools like LangChain \cite{langchain} or Guidance \cite{microsoftguidance}.
A significant advancement came with component-based frameworks \cite{angular, vuejs, react}, which encapsulate structure, style, and logic into reusable units, enhancing modularity and maintainability \cite{leff2001web, tatsubori2009html, vepslinen2023rise, vepslinen2023state}.
Building upon these frameworks, UI component libraries such as Ant Design \cite{antd} and MUI \cite{mui} provide extensive sets of pre-defined, high-level components (``\emph{widgets}'') that further accelerate development by offering ready-to-use solutions for common UI patterns, often including advanced data display capabilities.
Languages like JSX \cite{jsx}, used extensively in React \cite{react}, exemplify this by allowing developers to define UI ``\emph{components}'' using declarative, \emph{markup-like syntax} combined with programming logic.
This component-based approach improves code readability and fosters reuse.
Conversely, manually constructing structured outputs via string concatenation is notoriously error-prone and hinders readability \cite{carvalho2020text}.
These lessons underscore the potential benefits of applying similar principles --- using components, templates, and structured markup --- to manage the complexity of modern prompt engineering, a core motivation for POML.

\paragraph{Prompting with Markup Languages}
Several markup-based or component-based prompt frameworks have emerged, though they vary in their approach and capabilities.
Some approaches leverage JSX/TSX syntax within TypeScript (TS) for templating prompts \cite{vscodeprompttsx, priompt, prxmpt}.
These often prioritize adapting to context window lengths but may offer limited built-in features for complex data input handling or advanced styling mechanisms.
Other frameworks adopt broader React-inspired workflows or component models for orchestrating agent interactions or generating complex outputs \cite{aijsx, gensx}.
These systems typically offer limited control over the final prompt string's styling or the embedding of diverse data types directly via markup.
MDXPrompt \cite{mdxprompt} integrates MDX syntax for prompts but requires TypeScript API usage to render dynamic content, coupling the prompt definition to the execution environment.
These approaches attempt to ``componentize'' prompts with HTML-like syntax but are limited in ``widget-level'' data input encapsulation, or focus primarily on high-level agent or workflow orchestration rather than providing comprehensive prompt-level structuring and styling solutions.

Outside the JS/TS ecosystem, other markup solutions propose structured tags but typically focus more on conversational chat formats or specialized formats for LLM training \cite{chatxml, chatml}.
PromptML \cite{promptml}, while advocating for a domain-specific language to improve standardization and collaboration, currently lacks integrated development environment support, such as syntax highlighting, which can affect usability and adoption.
SAMMO \cite{schnabel2024symbolicpromptprogramsearch} operates at a higher-level abstraction, representing prompts symbolically to allow for transformations and optimization searches.
While these solutions provide intention-expressive tags or enable powerful symbolic manipulations, they generally lack built-in styling layers, flexible data import features across multiple formats, or comprehensive IDE toolkits.
Many existing markup or programmatic approaches are also tightly bound to specific programming languages (e.g., Python), limiting flexibility.
In contrast, POML aims to unify structured markup, effective data handling, decoupled styling, and integrated templating within a standalone language specification, supported by a comprehensive toolkit and SDKs for broader compatibility and ease of use across different development environments.
A table summarizing the key differences between POML and existing prompt markup languages is provided in Appendix \ref{sec:related_work_compare}.

\section{Motivations and Design Goals}

\subsection{Motivations}

Insights gathered from preliminary discussions with 5 prompt engineers in our research group highlighted 4 pressing challenges in contemporary prompt engineering.
These challenges directly motivate our design of POML as a comprehensive solution for prompt development.

\paragraph{M1: Structured Prompt Orchestration Necessity}
While the benefits of structured prompting are increasingly recognized \cite{schnabel2024prompts, promptml, chatxml, desmond2024exploring, wang2024langgpt, li2024guiding}, a significant problem lies in the lack of a standardized framework for implementation.
Current practices often result in ad-hoc approaches using unstructured plain text where instructions, roles, tasks, and examples are scattered inconsistently throughout prompts \cite{rtfprompt, chatgpt3freepromptlist, wang2024langgpt}.
This absence of standardization makes prompts difficult to maintain, optimize, and reuse, especially within collaborative development environments.

\paragraph{M2: Complexity in Data Integration}
Large LLM applications increasingly rely on varied data formats, requiring references to documents \cite{lin2024revolutionizingretrievalaugmentedgenerationenhanced}, tables \cite{zhang2024benchmarking, chen2024viseval}, or images \cite{liu2023visualinstructiontuning}.
Integrating these diverse data sources becomes chaotic and error-prone.
Manual text shaping, multi-point editing, and frequent reformatting significantly hamper maintainability \cite{desmond2024exploring, he2025tableloralowrankadaptationtable}.
Existing tools often rely on ad-hoc placeholders or separate scripts for data embedding \cite{semantickernel, langchain, bach2022promptsource}, lacking effective, integrated solutions.

\paragraph{M3: Prompt Styling and Format Sensitivity}
LLMs exhibit significant sensitivity to subtle variations in input formatting, where minor presentational changes can drastically alter results \cite{sclarquantifying, salinas2024butterfly, voronov2024mind, he2024does, zhuo2024prosa}.
This sensitivity underscores the need for structured control over prompt presentation.
The absence of a mechanism to decouple presentation style from prompt content makes it difficult to systematically test, refine, or adapt prompt formats for different models or tasks \cite{liu2025promptcontentenhancingllm, schnabel2024symbolicpromptprogramsearch}.

\paragraph{M4: Prompt Development and Management Challenges}
Current prompt development workflows suffer from poor diff visibility and version control \cite{desmond2024exploring}, making it difficult to track changes and collaborate effectively.
Existing prompt markup languages \cite{promptml,chatxml,mdxprompt} often lack syntax highlighting, auto-completion, preview and debugging tools, increasing the learning curve and error rate.
Many solutions \cite{vscodeprompttsx,gensx} are tightly bound to particular programming languages or frameworks, limiting their integration with existing development tools and workflows.
These limitations hinder the efficiency of development and make prompt management increasingly difficult as projects scale.

\subsection{Design Goals}

Based on these motivations, we established four core design goals for POML, each intended to address one of the identified challenges.
These goals focus on standardizing syntax, supporting effective data handling, separating styling from content, and providing developer-focused tooling.

\paragraph{DG1: Reusable and Maintainable Prompt Markup}
A primary goal is to provide a standardized, JSX-like markup language that inherently supports structured prompting practices (\textbf{M1}).
The language \emph{should} enhance prompt clarity and ensure that prompt logic is easily refactorable and shareable across projects or teams.
It \emph{should} support a hierarchical nesting structure to reflect logical relationships and improve flexibility.
Furthermore, the system \emph{should} offer commonly used elements like roles, tasks, or few-shot examples, promoting modularity and consistency in line with structured prompting approaches \cite{rtfprompt, wang2024langgpt}.

\paragraph{DG2: Comprehensive Data Handling}
To address the complexities of data integration (\textbf{M2}), the system \emph{should} offer effective and comprehensive data handling capabilities, including dynamic prompt generation.
Specialized components \emph{should} be provided to seamlessly embed commonly used static data sources into prompts, including documents, tables, and images \cite{chen2024viseval, xia2024agentless, lin2024revolutionizingretrievalaugmentedgenerationenhanced}.
These components \emph{should} preserve important data details while offering easy-to-use interfaces for adjusting presentation formats (e.g., syntaxes for tables) \cite{he2025tableloralowrankadaptationtable}, reducing the burden of manual text formatting.
Furthermore, the language itself \emph{should} support integrated templating logic (variables, loops, conditionals) as most data are only available at the runtime.
Support for fallback text for non-textual data \emph{should} also be included to accommodate LLMs with varying multi-modal capabilities and enhance prompt robustness across models.

\paragraph{DG3: Decoupled Presentation Specifications}
To manage LLM sensitivity to formatting (\textbf{M3}), a core design goal \emph{is} to provide flexible control over prompt presentation while simultaneously decoupling these presentation specifications (styling) from the core prompt content.
This separation \emph{should} allow styling rules (e.g., bullet style, syntax format, verbosity) to be defined centrally with minimized modifications to prompt contents.
The system \emph{should} support format configurations at multiple levels (e.g., both overall syntax and local syntax) to adapt prompts easily to different LLM sensitivities or task requirements \cite{he2024does, liu2025promptcontentenhancingllm}.
This decoupling will ideally enable rapid testing and optimization while maintaining content integrity.

\paragraph{DG4: Enhanced Tooling for Development}
To address development and management challenges (\textbf{M4}), the framework \emph{must} be accompanied by enhanced tooling.
It \emph{should} utilize standalone prompt files with a clear structure, facilitating integration with standard version control systems like Git and improving readability and collaborative editing.
Dedicated IDE support \emph{should} be provided with essential features like syntax highlighting, context-aware auto-completion, hover documentation, inline diagnostics, and a live preview.
Finally, the system \emph{should} maintain interoperability with popular LLM frameworks (e.g., LangChain \cite{langchain} and Guidance \cite{microsoftguidance}) through integration SDKs, allowing seamless integration into existing development workflows.

\section{POML: Prompt Orchestration Markup Language}
\label{sec:poml}

POML, standing for Prompt Orchestration Markup Language, introduces a novel structured paradigm designed for authoring, managing, and testing prompts for LLMs.
This section provides an overview of POML's foundational elements, including its core syntax based on HTML principles, its features for integrating diverse data types, its styling system for formatting prompt content, and a templating engine for dynamic prompt generation.
These features address common challenges in prompt engineering.
POML's hierarchical structure enhances clarity and reusability (\textbf{DG1}).
Its data components streamline the integration of varied information sources (\textbf{DG2}).
Its styling system manages LLM formatting sensitivities (\textbf{DG3}).
Its overall design combined with tooling (introduced in \S~\ref{sec:toolkit}) supports an enhanced development experience (\textbf{DG4}).

\subsection{Structured Prompting Markup}
\label{sec:poml_structure}

The fundamental syntax of POML adopts an HTML-like structure, closely mirroring HTML conventions.
This design choice stems from the observation that prompt engineering shares similarities with web design; both involve crafting interfaces --- textual for LLMs, visual for humans --- that require clear hierarchy, consistent presentation, and adaptability.
Using an HTML-inspired approach also leverages developers' familiarity with web technologies, reducing the learning curve \cite{promptdesign}.
Nested tags (called \textbf{components} in our case) form the overall structure, enabling the hierarchical composition of complex prompts from smaller, modular parts.
Within these tags, attributes provide metadata or configure the behavior of individual components, similar to how attributes like \texttt{src} or \texttt{href} function in HTML.
POML supports standard HTML comment syntax, \texttt{<!-- ... -->}, allowing developers to embed annotations within the prompt source code without affecting rendering.
POML also maintains compatibility with plain text, allowing for gradual adoption.
The root \texttt{<poml>} tag is optional, akin to the optional \texttt{<html>} tag in web development, which minimizes overhead for simple prompts where only specific features like data embedding might initially be needed.

\begin{figure}[t]
\includegraphics[width=\linewidth]{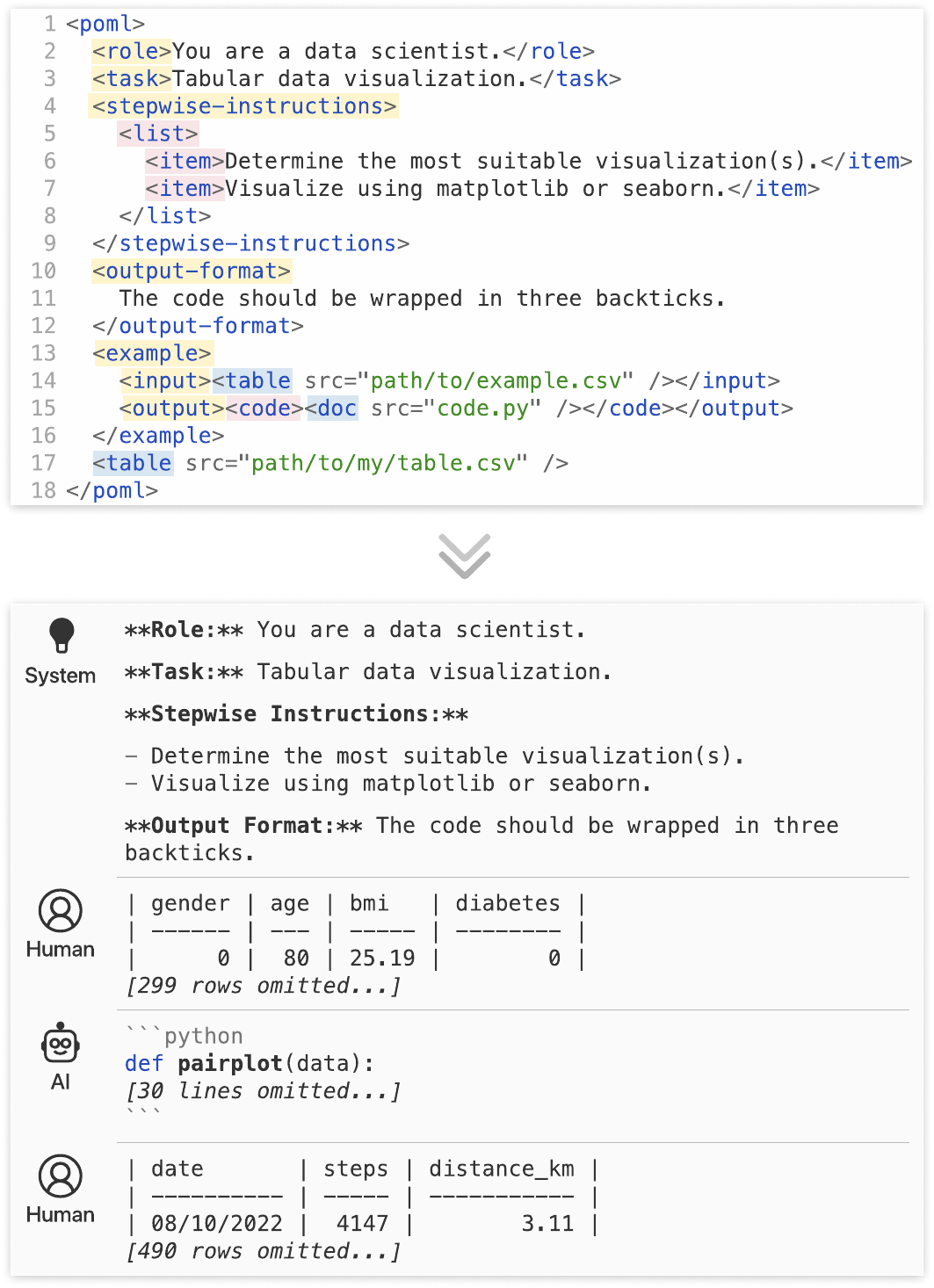}
\caption{
An example illustrating POML's structured markup and rendering (\S~\ref{sec:poml_structure}).
\textbf{Top:} POML source code demonstrating core components: intention components (e.g., \texttt{<role>}, \texttt{<task>}, \texttt{<example>}), data components (e.g., \texttt{<table>}, \texttt{<doc>}), and basic structural components (e.g., \texttt{<list>}, \texttt{<item>}).
\textbf{Bottom:} The corresponding rendered output, demonstrating how the structured markup translates into a clear prompt for an LLM. Note the rendering of intention components (e.g., \texttt{<role>} becomes the title ``\texttt{**Role:**}'') and data components (e.g., the first \texttt{<table>} is rendered as Markdown from its CSV source).
}
\label{fig:illustration-1}
\end{figure}

Functionally, POML components fall into several key categories, conceptually similar to how complex user interface widgets are built upon fundamental HTML elements.

\begin{itemize}
    \item \textbf{Basic Structural Components} provide fundamental text formatting and structural grouping capabilities, analogous to common HTML tags.
    Elements like \texttt{<b>} (bold), \texttt{<i>} (italic), \texttt{<p>} (paragraph), and \texttt{<div>} (division) allow for control over text presentation and logical grouping within larger content blocks, enhancing readability and structure.
    Alongside these static elements, POML includes components and syntax for dynamic content generation through its integrated templating engine, such as constructs for loops (\texttt{for}), conditionals (\texttt{if}), variable definition (\texttt{<let>}), and substitution (\texttt{{\{\{}...{\}\}}}).
    These templating features are detailed further in \S~\ref{sec:poml_templating}.

    \item \textbf{Intention Components} define the core logic and overall structure of the prompt.
    These components implement the ``structured prompting'' paradigm \cite{mdxprompt,promptml,wang2024langgpt,chatgpt3freepromptlist,rtfprompt}, establishing the interaction's purpose and guiding the LLM's response.
    Examples include the \texttt{<role>} component, used to define the persona the LLM should adopt; the \texttt{<task>} component, which specifies the objective the LLM needs to achieve; and the \texttt{<example>} component, crucial for providing few-shot demonstrations to guide the model's behavior.
    Unlike data components, intention components organize the logical blocks of the prompt rather than presenting complex data directly.
    When rendered, components like \texttt{<role>} typically appear as formatted titles (e.g., \texttt{**Role:**}), with the presentation adjustable via styling rules (\S~\ref{sec:poml_styles}).
    These intention components orchestrate the overall flow and content of the prompt, as illustrated in \autoref{fig:illustration-1}.

    \item \textbf{Data Components} are designed to handle the integration of diverse external data formats into the prompt context.
    Elements such as \texttt{<document>}, \texttt{<img>}, and \texttt{<table>} provide methods to embed content from files or data structures.
    These components, detailed in \S~\ref{sec:poml_data_components}, are essential for grounding LLM responses in specific information or enabling tasks that operate on external data.
\end{itemize}

\paragraph{Rationale}
This hierarchical and descriptive markup enforces a logical organization, offering significant advantages over unstructured plain text.
For instance, as illustrated in \autoref{fig:illustration-1}, the explicit separation of conceptual parts --- using intention components like \texttt{<role>} and \texttt{<task>}, structuring instructions with \texttt{<stepwise-instructions>}, defining outputs with \texttt{<output-format>}, and providing few-shot demonstrations \cite{brown2020language} via \texttt{<example>} (which itself nests \texttt{<input>} containing data components like \texttt{<table>} and \texttt{<output>} referencing an external \texttt{<doc>}) --- significantly improves prompt clarity and understandability, reducing ambiguity for both LLMs and human readers.
These components, such as the tables shown integrated directly or within examples, can then be easily reused across different prompts or modified in isolation, fostering systematic prompt engineering (\textbf{DG1}).
Compared to simpler formats like PromptML \cite{promptml}, which primarily offer basic task and example definitions, POML provides richer semantic components crucial for handling diverse content types (\textbf{DG2}).
Meanwhile, the modularity allows individual components to be treated as self-contained units.
Modifying components independently minimizes the risk of unintended side effects elsewhere in the prompt, facilitating faster and safer iterations, supporting more agile development workflows.
The combination of a standardized, text-based format and clearly named elements simplifies prompt management within standard version control systems like Git (\autoref{fig:teaser} (a)).
This enhances collaboration, as changes become easier to track, merge, and discuss (\textbf{DG4}).

\subsection{Data Components} 
\label{sec:poml_data_components}

\begin{figure*}
  \includegraphics[width=\textwidth]{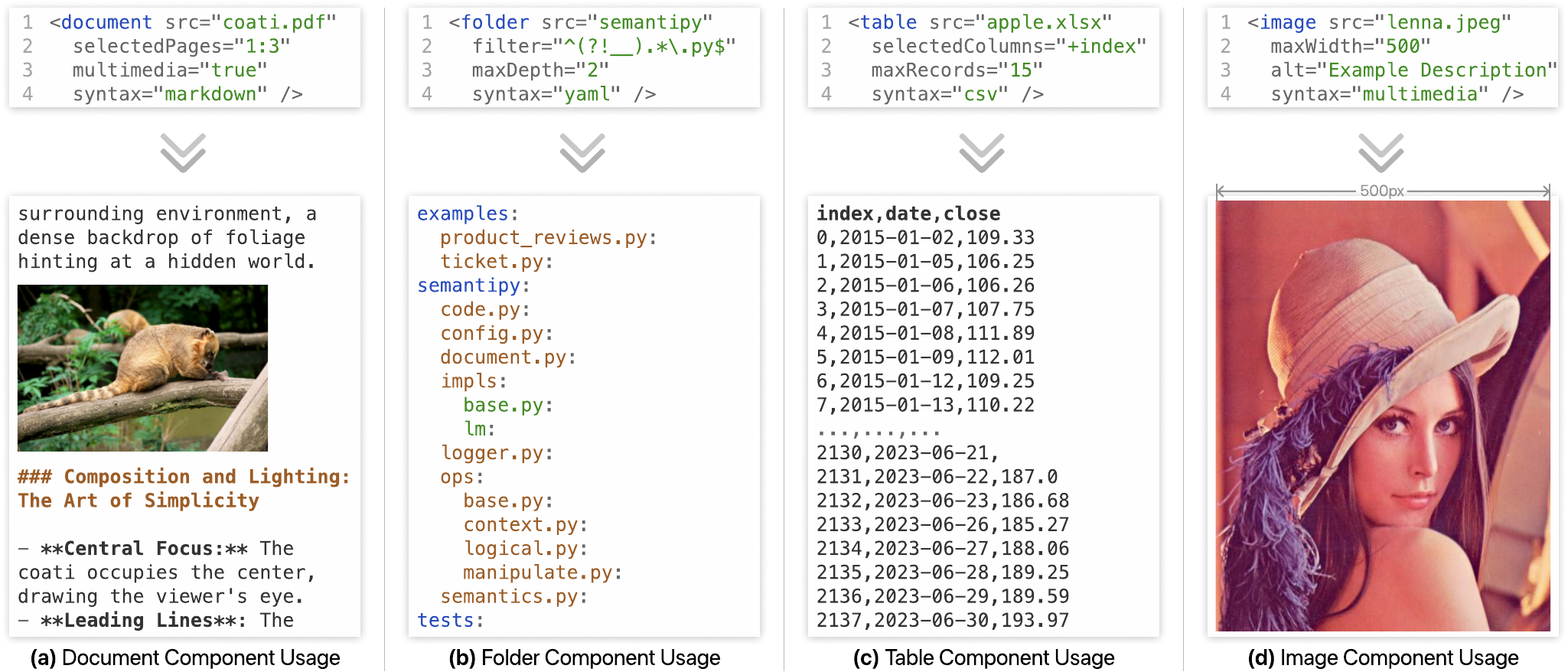}
  \caption{Examples of POML data components demonstrating integration of diverse data types (\S~\ref{sec:poml_data_components}). \textbf{(a)} \texttt{<document>} rendering selected PDF pages with multimedia; \textbf{(b)} \texttt{<folder>} displaying a filtered directory structure as YAML; \textbf{(c)} \texttt{<table>} extracting and formatting spreadsheet data as CSV; \textbf{(d)} \texttt{<img>} inserting a referenced image with resizing.}
  \label{fig:data-components}
\end{figure*}

POML provides specialized data components to systematically integrate diverse data modalities directly within the prompt structure, as exemplified in \autoref{fig:data-components}.
This capability directly addresses the critical need to combine LLM reasoning with varied external data sources, a key requirement for many advanced applications (\textbf{DG2}).
POML's modular data components reduce the inherent complexity associated with constructing prompts that incorporate multiple data inputs.
These built-in components distinguish POML by offering systematic handling for common data types within a unified framework, analogous to component libraries in modern UI frameworks \cite{mui,antd}.
Currently, POML supports \textbf{7 distinct data components}: document, table, folder, image, conversational messages, audio clip, web page.
We detail \textbf{4 representative examples} here; the remaining components operate similarly and are demonstrated in the case studies (\S~\ref{sec:case_studies}).

\paragraph{Document}
The document component (\autoref{fig:data-components} (a)) provides a mechanism for referencing and embedding content from external text-based files, supporting formats like \texttt{.txt}, \texttt{.docx}, and \texttt{.pdf}.
This is particularly crucial for tasks involving large amounts of background information or context, such as in retrieval-augmented generation (RAG) systems \cite{lin2024revolutionizingretrievalaugmentedgenerationenhanced,chatpdf}.
It offers granular control through attributes, allowing developers to specify character encoding, whether to preserve original formatting, whether to include or discard embedded images within documents like PDFs, and how to trim content --- for instance, by selecting specific page ranges (e.g., \texttt{selectedPages="1:3"}) --- thereby helping manage the LLM's limited context window.

\paragraph{Folder}
For scenarios involving interaction with file systems or code repositories, such as file system analysis, code reviews, and project navigation \cite{jimenez2024swebench,liu2023agentbench,uithub}, the \texttt{<folder>} component (\autoref{fig:data-components} (b)) provides a structured way to represent directory hierarchies.
It allows customization of the representation through attributes controlling the maximum depth of the displayed tree, filtering options to include or exclude files based on extensions or name patterns (e.g., \texttt{filter="\^{}(?!\_\_).*\textbackslash.py"}), and optional automatic summarization for very large directories to conserve context space.
It can also display metadata like file sizes or modification dates and supports multiple output formats, including classic tree views using box-drawing characters, YAML, or JSON representations.

\paragraph{Table}
Tables are vital for complex question-answering over structured data \cite{pasupat2015compositional}, generating data visualizations \cite{chen2024viseval}, or performing data manipulation tasks \cite{zhang2024benchmarking}.
The \texttt{<table>} component (\autoref{fig:data-components} (c)) supports a variety of input sources, including CSV, TSV, Excel spreadsheets, and JSON arrays of objects.
Crucially, it allows specifying the output representation (e.g., Markdown, HTML, XML, plain CSV) and controlling content details such as the inclusion of headers or indices, and selective presentation of rows or columns for large tables, offering flexibility in presentation.
Employing a consistent and well-defined table format via this component can significantly enhance an LLM's ability to accurately parse and reason over the tabular data, applicable to scenarios like \cite{he2025tableloralowrankadaptationtable,sui2024tablemeetsllmlarge,zhang2024benchmarking}.

\paragraph{Image}
Acknowledging the rapid advancement of multimodal LLMs capable of processing visual information \cite{geminiteam2024gemini15unlockingmultimodal,openai2024gpt4technicalreport}, POML includes the \texttt{<img>} component (\autoref{fig:data-components} (d)) for direct image insertion into prompts.
Inspired by web accessibility principles \cite{w3c, w3caccessibility, w3caccessibilityprinciples}, this component supports standard attributes like \texttt{alt} text, allowing prompts to be easily adapted to LLMs with varying capabilities, including those without vision, thereby enhancing robustness.
Layout can be influenced using a \texttt{position} attribute (e.g., \texttt{before}, \texttt{after}, \texttt{here}) to control placement relative to surrounding text.
Attributes like \texttt{maxWidth} and \texttt{maxHeight} allow developers to suggest rendering constraints, potentially influencing the number of tokens consumed when sending image data to the model.
Compared to interactive uploads in chat interfaces \cite{openaichatgpt,gemini2} or less structured parameter passing in raw API calls \cite{geminiteam2024gemini15unlockingmultimodal,openai2024gpt4technicalreport}, inserting images via a dedicated component within the markup offers superior programmatic control and integration.

\paragraph{Extensions} POML's architecture is designed to be extensible.
As LLMs evolve to handle new modalities, POML can incorporate additional data components for types like video segments and node-edge graphs, maintaining a consistent framework for diverse data integration.

\subsection{Styling System}
\label{sec:poml_styles}

\begin{figure*}
  \includegraphics[width=\textwidth]{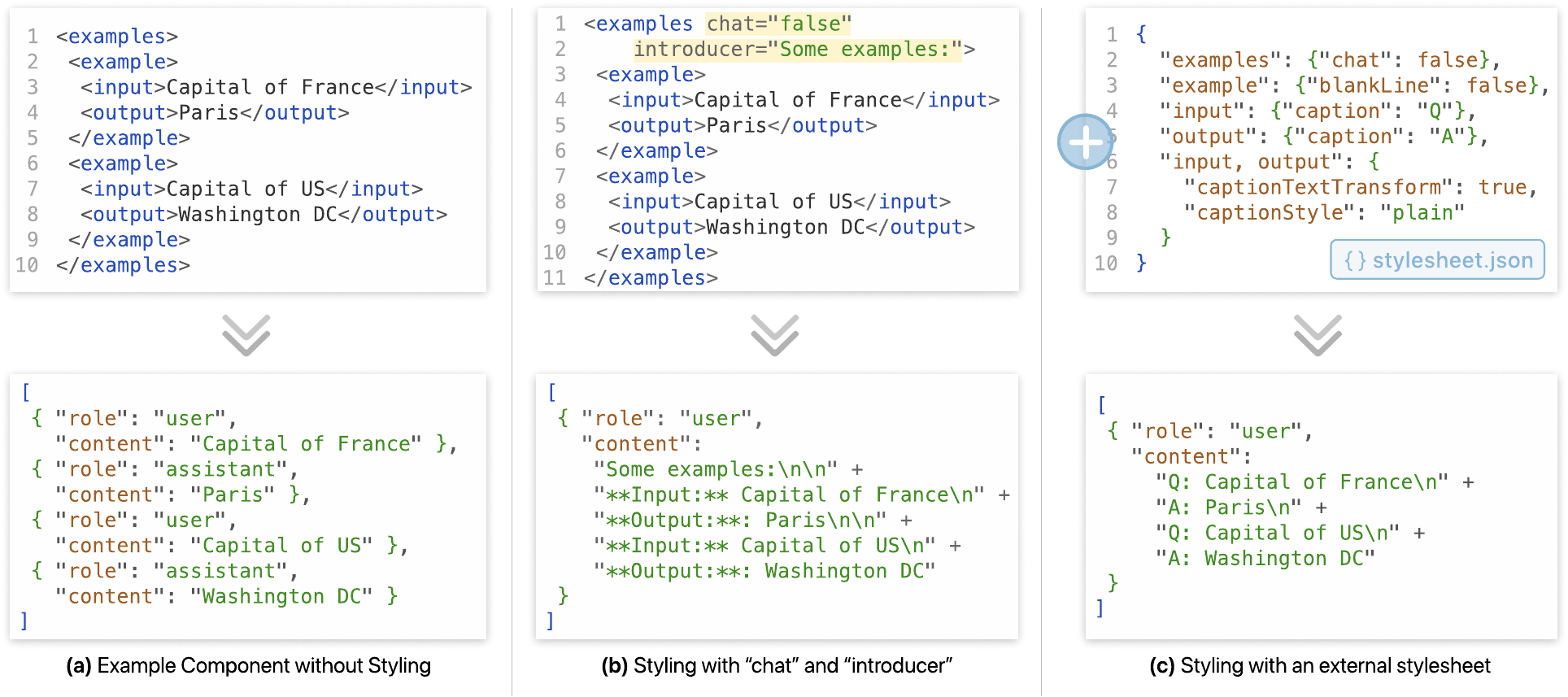}
  \caption{Demonstrating POML styling capabilities (\S~\ref{sec:poml_styles}). \textbf{(a)} Default rendering of \texttt{<example>} components. \textbf{(b)} Inline \texttt{chat} and \texttt{introducer} attributes on the parent \texttt{<examples>} element modify its presentation (from chat messages to plain text with ``\texttt{**Input**}''/``\texttt{**Output**}'' captions). \textbf{(c)} A \texttt{<stylesheet>} applies global rules to POML in (a), controlling layout (\texttt{chat=false}), captions/prefixes (\texttt{caption="Q:"}/\texttt{"A:"}), and styles (\texttt{captionStyle="plain"}), resulting in customized output.}
  \label{fig:style-components}
\end{figure*}

A critical challenge in prompt engineering stems from the documented sensitivity of LLMs to subtle variations in input formatting \cite{sclarquantifying, salinas2024butterfly, voronov2024mind, he2024does, zhuo2024prosa}.
Minor changes in presentation can significantly impact model performance, necessitating precise control over how prompt content is rendered (\textbf{DG3}).
POML addresses this through a dedicated styling system designed to manage presentation effectively.
The styling options provided, such as control over syntax formats, layout adjustments (e.g., chat vs. block formatting), list styles, and caption presentation, are informed by findings in prompt engineering research exploring these sensitivities \cite{liu2025promptcontentenhancingllm, sclarquantifying}.
Inspired by the separation of concerns in web development (HTML for structure, CSS for style \cite{css, w3c}), POML deliberately decouples presentation rules from the core prompt content.
POML offers two primary mechanisms for applying styles: \emph{inline attributes} and \emph{external stylesheets}.

POML provides control over various presentation aspects known to influence LLM behavior, such as layout adjustments (e.g., chat versus block formatting), list styles (\texttt{<list list\-Style="deci\-mal">}), caption presentation (\texttt{<hint caption\-="\-Import\-ant \-Note" captionStyle\-="header">}), and overall verbosity \cite{liu2025promptcontentenhancingllm, sclarquantifying}.
Beyond simple formatting, POML provides a crucial \textbf{\texttt{syntax} attribute}.
This attribute allows developers to explicitly control the rendering format for specific components or the entire prompt (e.g., \texttt{<table syntax="html"/>} or \texttt{<poml syntax="json">}).
This control is vital because different LLMs can exhibit varying sensitivities or capabilities when parsing structured formats like Markdown, JSON, or XML \cite{claudexml} embedded within the prompt text.
The styling attributes can be nested, allowing, for example, a subsection of a predominantly Markdown-formatted prompt to be rendered as JSON, providing fine-grained control over the final string sent to the model.
One way to apply these styling adjustments is via \textbf{inline attributes} on specific POML components --- a convenient method for localized formatting familiar to web developers.
\autoref{fig:style-components} ((b) compared to (a)) demonstrates how inline attributes can alter the presentation of \texttt{<example>} components.

For managing styles more systematically across multiple components or prompts, POML supports \textbf{external stylesheets}, typically defined in separate JSON files (e.g., \texttt{stylesheet.json} as shown in \autoref{fig:style-components} (c)).
These files contain global formatting rules defined using a concise, JSON-like syntax.
Style rules within the stylesheet target specific POML component types (e.g., \texttt{hint}, \texttt{table}) or user-defined classes (see \S~\ref{sec:cs_tableqa} for examples).
Alternatively, these style rules can also be embedded directly within a POML document using the \texttt{<stylesheet>} tag, as shown in \autoref{fig:teaser} (c).
Overall, stylesheets provide a mechanism to define styles that apply globally or to specific component types, offering a way to batch-manage or customize styles that might otherwise be set inline individually.
They offer several advantages: (1) it provides centralized control, establishing a single source of truth for formatting; (2) it keeps the primary prompt logic cleaner by avoiding repetitive presentation attributes; (3) it simplifies style modifications by eliminating multi-point edits, enhancing maintainability.
As a result, this approach directly facilitates systematic experimentation with different presentation formats to address LLM sensitivities or task requirements (\textbf{DG3}) without altering the core prompt content.

\subsection{Templating Engine}
\label{sec:poml_templating}

\begin{figure}[t]
  \includegraphics[width=\linewidth]{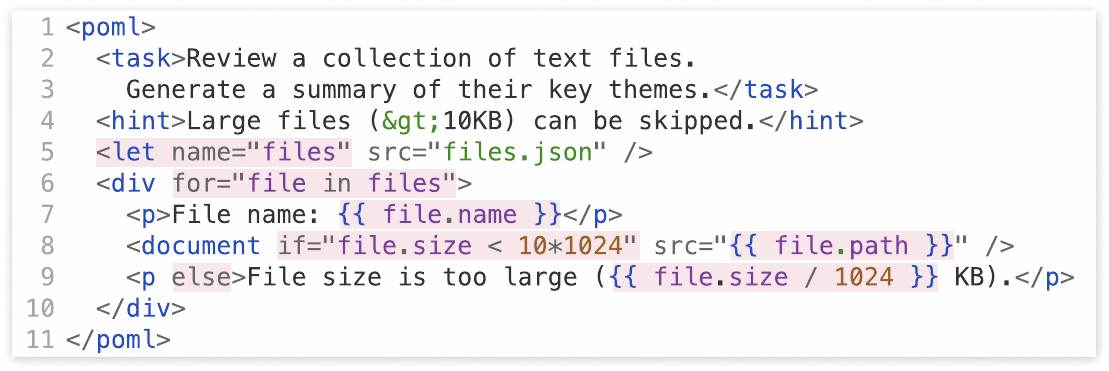}
  \caption{Example of POML's templating engine (\S~\ref{sec:poml_templating}): using \texttt{<let>} to load data (from \texttt{files.json}), \texttt{for} attribute to iterate over items, \texttt{{\{\{} ... {\}\}}} for variable substitution (e.g., \texttt{{\{\{} file.name {\}\}}}), and \texttt{if}/\texttt{else} attributes for conditional component rendering based on data values (embedding a \texttt{<document>} only if \texttt{file.size} is below a threshold)}
  \label{fig:illustration-2}
\end{figure}

POML integrates a built-in templating engine to facilitate the creation of dynamic, data-driven prompts without execution environment dependencies (\textbf{DG2}).
Inspired by established web templating systems \cite{django, jinja, handlebars, jsx}, this engine allows runtime customization without relying solely on external scripting.
Key features include variable substitution using \texttt{{\{\{}variable{\}\}}} syntax, iteration over data collections via the \texttt{for} attribute, conditional rendering using \texttt{if}/\texttt{else}-like attributes, and inline variable definition with the \texttt{<let>} tag.
\autoref{fig:illustration-2} demonstrates these capabilities, showing how data loaded via \texttt{<let>} is iterated over with \texttt{for}, variables like \texttt{{\{\{} file.name {\}\}}} are substituted, and component rendering is controlled conditionally based on \texttt{file.size}.

This integrated approach offers advantages over manual string manipulation, which is often error-prone and hard to debug \cite{carvalho2020text}.
It reduces redundancy by enabling reusable prompt structures populated with varying data.
The declarative nature enhances readability, providing a clearer view of the prompt structure.
Furthermore, the engine is independent from other programming languages, differentiating POML from other solutions relying on existing templating engines (e.g., \cite{langchain}), contributing to a cohesive framework where structure, presentation, data integration, and dynamic logic are managed within a single language.
Details on the templating engine's syntax and capabilities are available in Appendix \ref{sec:poml_templating_detail}.

\section{Development Toolkit}
\label{sec:toolkit}

Complementing the core POML language specification, a comprehensive development toolkit is provided to enhance the efficiency, reliability, and collaborative nature of the prompt engineering process, directly addressing (\textbf{DG4}).
While POML's architecture allows potential integration with specialized prompt engineering environments like PromptIDE \cite{strobelt2022interactive} or ChainForge \cite{arawjo2024chainforge}, our initial implementation focuses on providing a feature-rich experience within Visual Studio Code, a widely adopted and familiar IDE.
This toolkit encompasses two key aspects: (1) integrated POML IntelliSense for rapid issue detection and code authoring assistance, and (2) Software Development Kits (SDKs) enabling integration of POML into established software development ecosystems.

\subsection{POML IntelliSense}
\label{sec:diagnostics}

POML IntelliSense refers to a suite of integrated features designed to simplify the prompt authoring process within the IDE, providing real-time feedback and assistance.

\paragraph{Syntax Highlighting}
Syntax highlighting, adapted from standard HTML conventions due to POML's similar structure, visually differentiates component tags, attributes, and content.
This improves code readability and aids in identifying structural elements.

\begin{figure}[t]
  \includegraphics[width=\linewidth]{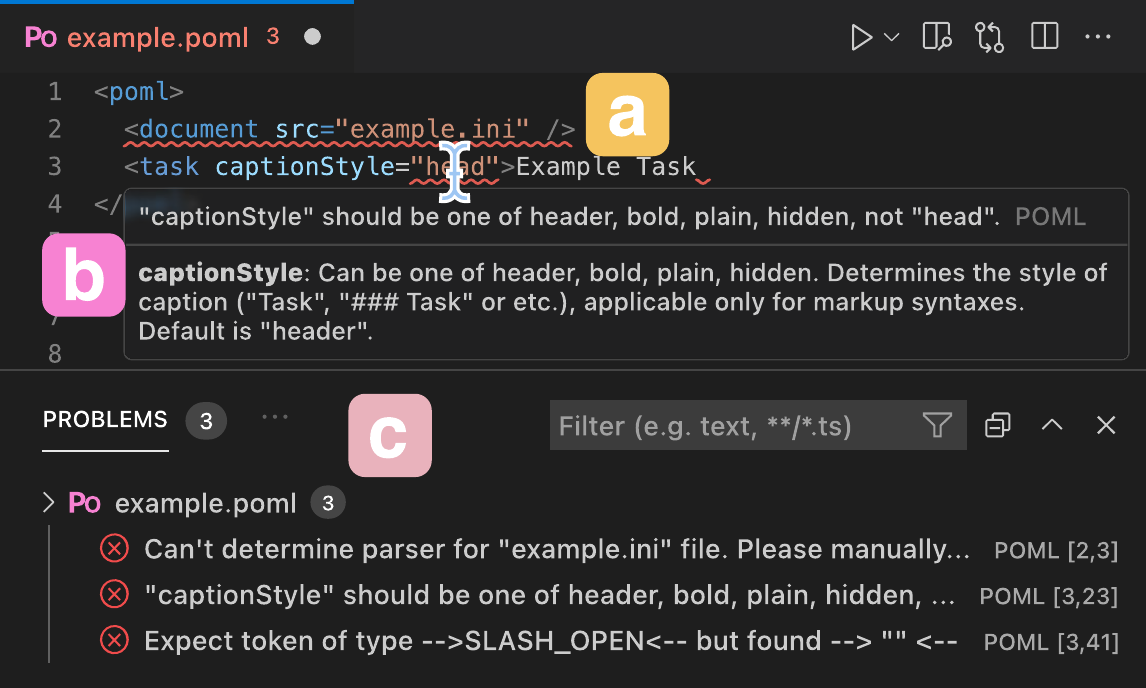}
  \caption{Real-time diagnostic feedback from the POML toolkit in the IDE.
  \textbf{(a)} Errors indicated directly in the editor via inline highlighting (e.g., missing attribute).
  \textbf{(b)} Hovering over an error shows detailed validation messages and documentation.
  \textbf{(c)} A dedicated panel lists all detected issues (syntax errors, validation problems, file processing issues).}
  \label{fig:diagnosis}
\end{figure}

\paragraph{Hover and Auto Completion}
Informative hover tooltips and context-aware auto-completion enhance productivity.
Hovering over POML elements displays dynamically retrieved documentation, definitions, and examples (\autoref{fig:diagnosis} (b); \autoref{fig:teaser} (d)), ensuring up-to-date information.
As the developer types, the system provides context-aware auto-completion suggestions (\autoref{fig:teaser} (f)).
It intelligently suggests valid POML component names based on the current nesting context, available attributes for a given component, and permissible values for certain attributes (e.g., predefined style options), reducing errors and accelerating development by facilitating syntax discovery.

\paragraph{Inline Diagnostics}
Real-time inline diagnostics provide immediate feedback within the editor (\autoref{fig:diagnosis}; \autoref{fig:teaser} (e)).
Potential syntax errors, invalid attributes, or structural issues are automatically detected and highlighted with descriptive messages attached to the problematic lines for rapid identification and correction.
The system employs error tolerance, attempting to recover from detected issues and reporting multiple issues simultaneously (\autoref{fig:diagnosis} (c)), allowing developers to address several problems efficiently.

\begin{figure}[t]
  \includegraphics[width=\linewidth]{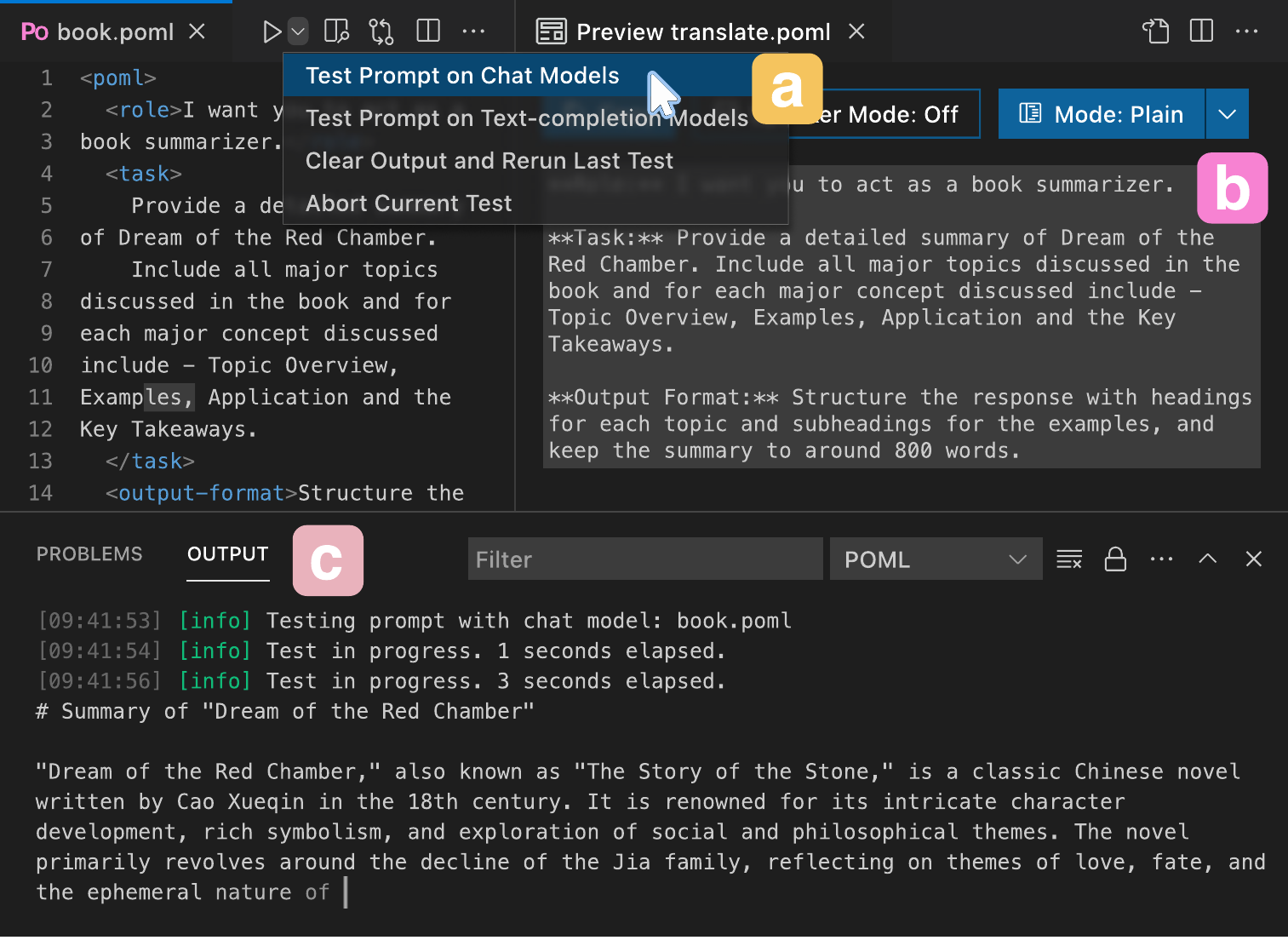}
  \caption{Integrated interactive prompt testing within the POML development environment.
  \textbf{(a)} Users initiate or abort tests against specific model types (e.g., chat or text-completion) via a context menu directly from the editor.
  \textbf{(b)} The live preview panel displays the prompt being tested.
  \textbf{(c)} The output panel shows test progress logs and streams the LLM's response in real-time.}
  \label{fig:interactive-testing}
\end{figure}

\paragraph{Live Preview}
Working in concert with inline error reporting is a dynamic live preview feature (\autoref{fig:interactive-testing} (b); \autoref{fig:teaser} (g)).
This feature renders a visual representation of the final prompt structure as the POML code is being written, updating automatically with each change.
This immediate visual feedback helps verify component hierarchy, content, styling, and templating, reducing the cognitive load \cite{priompt} of mentally translating markup.
The preview mechanism supports multiple viewing modes tailored to different needs: a ``speaker mode'' for chat-based LLMs, a ``text mode'' for text completion models, a visually enhanced ``render mode'' for readability, and a ``plain text mode'' showing the exact raw string for debugging subtle spacing issues.

\paragraph{Interactive Testing}
Integrated interactive testing allows developers to initiate LLM tests directly within the editor (\autoref{fig:interactive-testing}).
Users can select model types (e.g., chat, text-completion) via a context menu (\autoref{fig:interactive-testing} (a)) and test the current prompt without context switching.
The output panel (\autoref{fig:interactive-testing} (c)) displays test progress logs and streams the LLM's response in real-time, enabling immediate observation and early abortion of the generation in case it starts to deviate from the desired outcome (\autoref{fig:interactive-testing} (a)).
This tight feedback loop facilitates iterative refinement and assessment of prompt compatibility across LLMs, addressing model sensitivities (\textbf{DG3}) and accelerating the write-debug-validate cycle.

\paragraph{LSP Server}
The IntelliSense features described above --- including hover information, auto-completion, and inline diagnostics --- are powered by a dedicated implementation of the Language Server Protocol (LSP) \cite{lsp}.
Leveraging the standardized LSP allows these rich language-specific capabilities to be provided consistently.
Caching and throttling are implemented at the side of LSP server, ensuring that the client (e.g., VSCode) remains responsive and does not experience performance degradation during heavy usage.
This architectural choice not only enhances the development experience currently within VSCode but also creates the potential for integrating POML IntelliSense into other LSP-compatible editors (e.g., Emacs, PyCharm) in the future.

\subsection{Node.js and Python SDKs}
\label{sec:sdks}

\begin{figure}[t]
  \includegraphics[width=\linewidth]{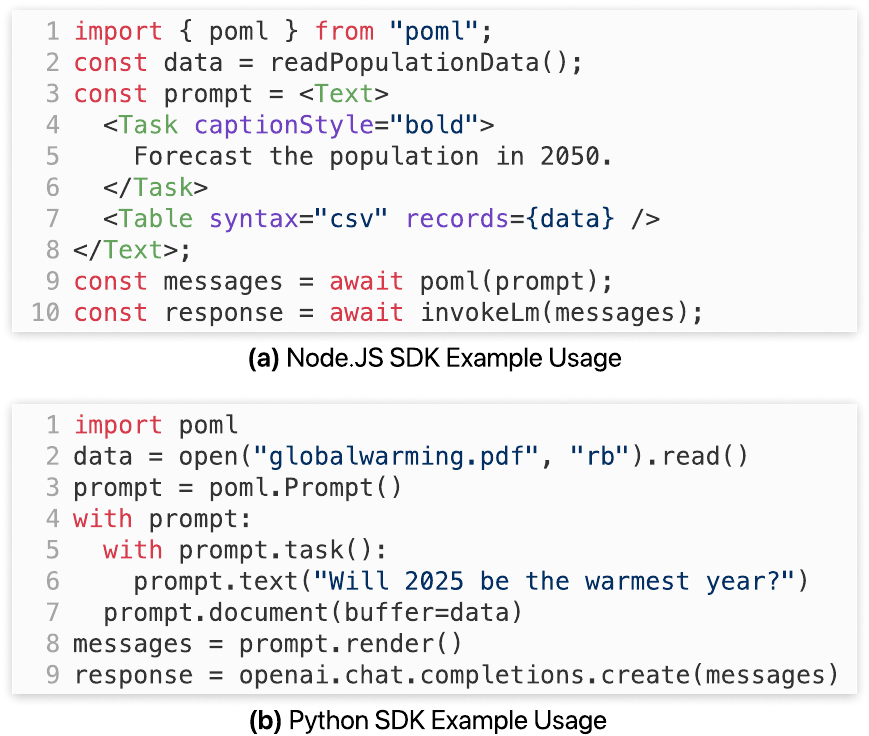}
  \caption{Using POML Software Development Kits (SDKs) to integrate into programming workflows. \textbf{(a)} JavaScript/TypeScript SDK example using JSX-like tagged template literals. \textbf{(b)} Python SDK example using a context manager approach.}
  \label{fig:sdks}
\end{figure}

To facilitate integration into larger software applications, POML provides Software Development Kits (SDKs) for the Node.js (JavaScript / TypeScript) and Python environments, chosen for their prevalence in web development and AI respectively (\autoref{fig:sdks}).

The \textbf{Node.js SDK} (\autoref{fig:sdks} (a)) allows programmatic POML definition using familiar JavaScript paradigms like JSX syntax or React-style functional components \cite{vscodeprompttsx, gensx}.
This approach supports creating typed, composable prompt components that can be versioned and shared using standard tools like npm.
It also enables developers to define custom, extensible POML components using TypeScript or JavaScript, allowing the core language to be augmented within specific application contexts.

Similarly, the \textbf{Python SDK} (\autoref{fig:sdks} (b)) offers an idiomatic interface using context managers and a fluent API builder pattern, inspired by libraries like \texttt{yattag} \cite{yattag}.
This design facilitates integration into Python-based AI/ML workflows, for instance, dynamically creating prompts within data processing pipelines or interacting with frameworks like LangChain \cite{langchain}.
Importantly, the Python tooling also supports a standalone mode, allowing direct processing of \texttt{.poml} files via scripts or command-line tools, ideal for rapid testing as demonstrated in our TableQA case study (\S~\ref{sec:cs_tableqa}).

Crucially, both SDKs support programmatic generation and manipulation of stylesheets.
This capability enables advanced use cases, such as the automated exploration and optimization of prompt styles.
They are also compatible with popular LLM client libraries, simplifying the integration of POML-generated prompts into the LLM communication layer.
Furthermore, the language-agnostic design of the core POML specification also allows for the potential development of SDKs in other languages (e.g., C\#, Java, Go) in the future.

\section{Implementation}
\label{sec:implementation}

\paragraph{Core Technology Stack}
The core POML engine is developed using TypeScript, leveraging the React framework combined with Server-Side Rendering (SSR) \cite{servercomponents}.
React's component model naturally supports POML's design philosophy centered on separable and reusable prompt components.
SSR facilitates the efficient handling of asynchronous operations, such as processing external data sources like images or documents referenced within prompts.

\paragraph{Codebase Overview}
The POML codebase currently comprises approximately 14.8k lines of TypeScript code.
This implementation realizes the POML language specification, encompassing 37 distinct components (categorized into 14 basic structural, 7 data, 11 intention, and 5 utility components) and a total of 283 attributes.
The codebase is divided into approximately 11.3k lines for the core engine functionality, 3.4k lines for the Visual Studio Code extension features (\S~\ref{sec:diagnostics}), and 96 lines for a lightweight Python SDK wrapper (\S~\ref{sec:sdks}).
Comprehensive testing validates the implementation through 10 test suites containing 115 distinct test cases.
These tests cover diverse scenarios, including basic tag parsing, attribute validation, complex multi-modal data integration, nested component structures, and the correct application of styling rules, verifying robustness of core POML features.

\paragraph{Rendering Architecture}
The POML implementation employs a three-pass rendering architecture to enhance modularity and extensibility.
First, a \textbf{Parser} pass validates the input POML markup, executes templating logic (loops, conditionals), resolves variables, applies styles, and transforms the source code into React JSX components.
Second, \textbf{React} processes these components, and generates a comprehensive \textbf{Intermediate Representation (IR)}.
This IR is a structured tree containing the resolved content, computed styles, and associated metadata.
Third, a dedicated \textbf{Writer} pass traverses the IR and serializes it into the final target output format (e.g., Markdown, JSON, plain text).
This separation of concerns significantly enhances flexibility; for instance, it enables support for new output formats by implementing additional Writers without modifying the core parsing or IR generation stages.
Furthermore, this architecture potentially allows support for alternative input syntaxes beyond XML, such as Markdown extensions or YAML, by developing corresponding Parser modules.
It also facilitates performance optimizations like IR caching.
A detailed explanation of the rendering architecture is available in Appendix~\ref{sec:three_pass_rendering}.

\section{Case Studies}
\label{sec:case_studies}

\subsection{PomLink -- Prototype of iOS Agent}
\label{sec:cs_pomlink}

\begin{figure}
  \centering
  \begin{minipage}[b]{0.49\linewidth}
    \centering
    \includegraphics[width=0.95\linewidth]{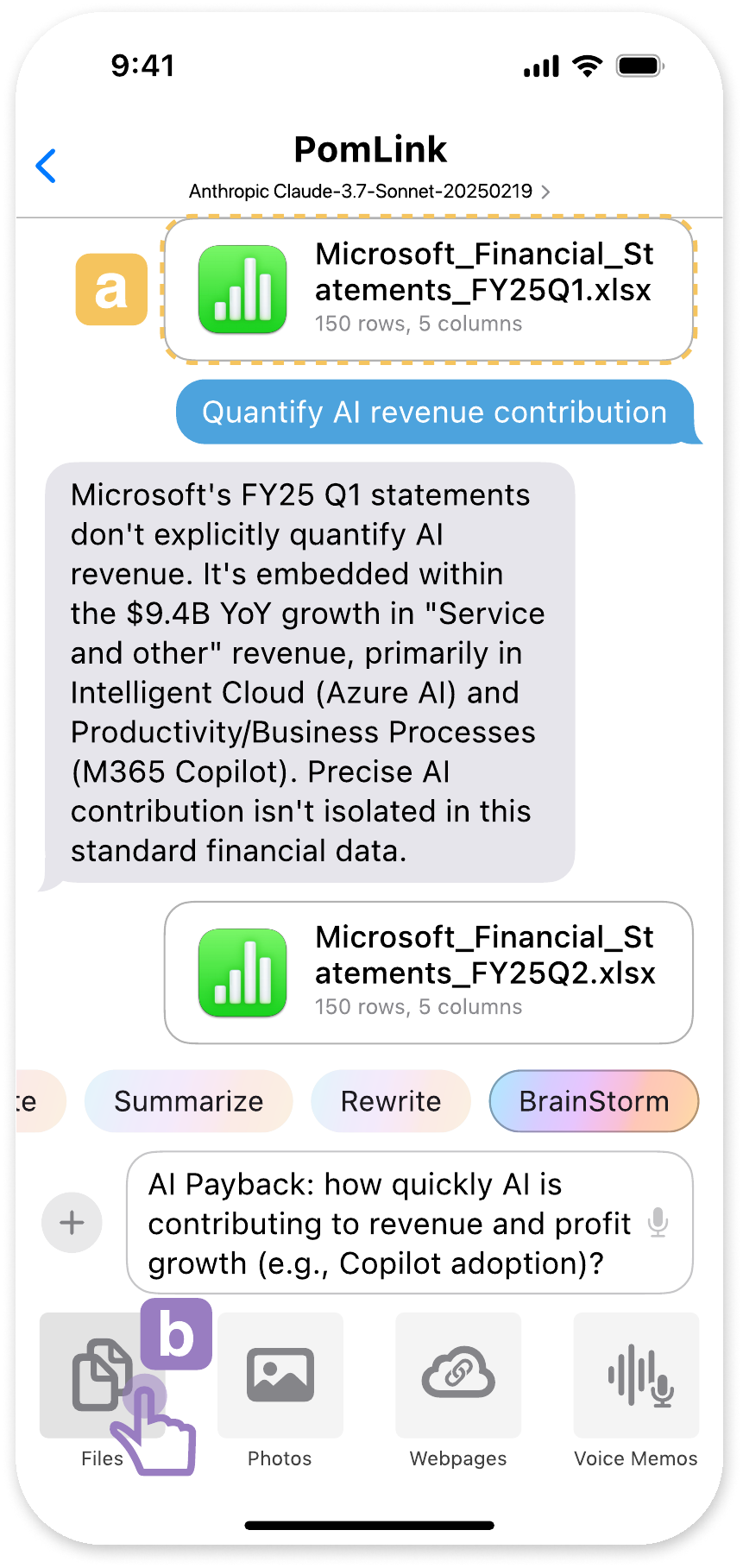}
    \vspace{-3mm}
    \caption*{(1)}
  \end{minipage}%
  \begin{minipage}[b]{0.49\linewidth}
    \centering
    \includegraphics[width=0.95\linewidth]{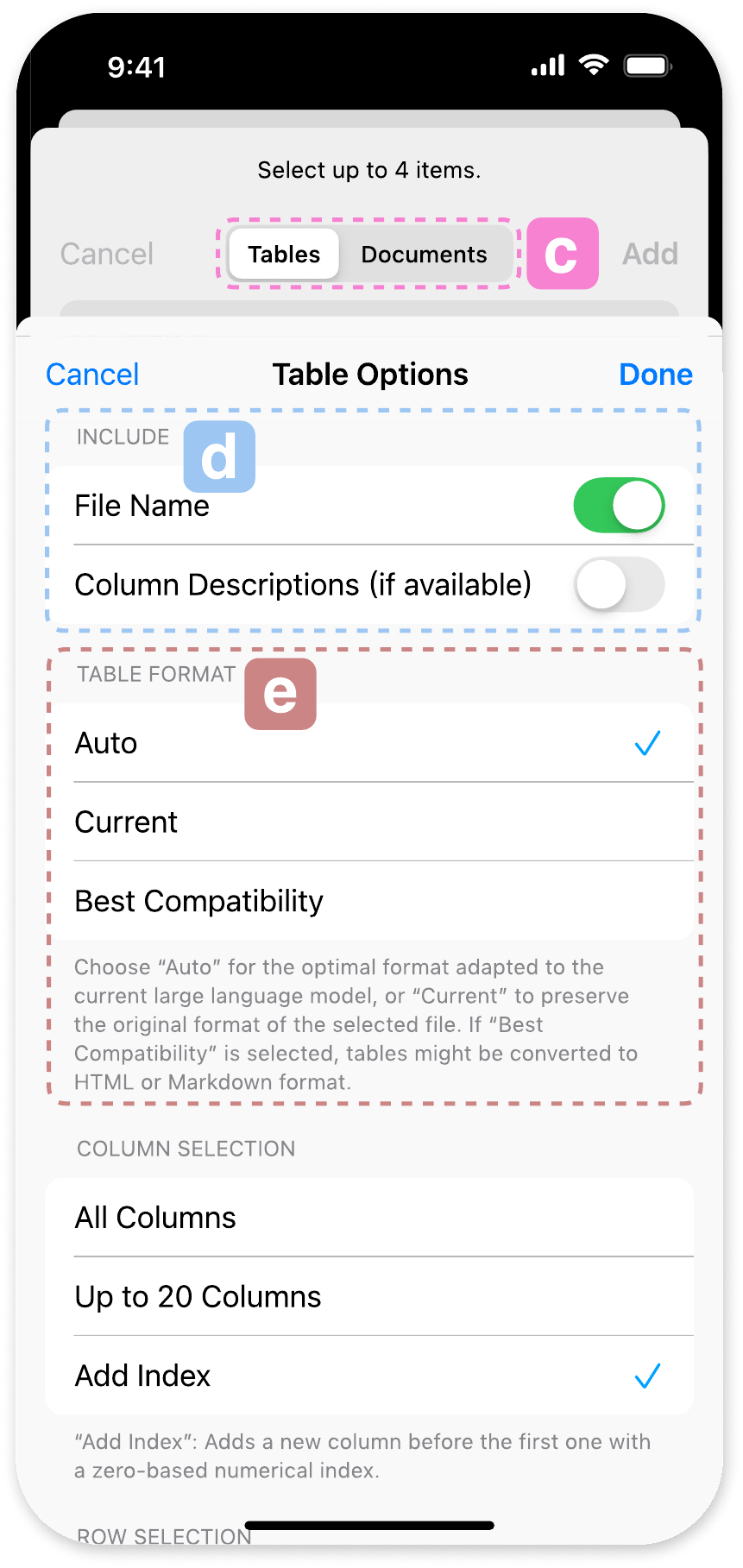}
    \vspace{-3mm}
    \caption*{(2)}
  \end{minipage}
  \vspace{-1mm}
  \caption{PomLink iOS interface, powered by POML.
  \textbf{(1)} Main chat screen showing LLM interaction: \textbf{(a)} linked Excel data within the conversation; \textbf{(b)} buttons for adding various file types.
  \textbf{(2)} Table configuration panel, derived from POML's \texttt{<table>} component features, where users can select \textbf{(c)} documents/tables, configure \textbf{(d)} inclusion options (e.g., file name), and choose \textbf{(e)} table formatting preferences (including an ``Auto'' option)}
  \label{fig:pomlink-table-selector}
\end{figure}

To demonstrate POML's utility in developing complex, data-intensive applications, we created PomLink, an iOS agent prototype.
This case study highlights POML's capability to streamline the integration of diverse data sources within a functional LLM-powered mobile application.
PomLink functions as an LLM agent that utilizes context derived from ``\emph{linked}'' files, which users can upload via dedicated interface buttons supporting various types including documents, tables, images, webpages, and voice memos (\autoref{fig:pomlink-table-selector} (1)), all of which can be tailored to personalized settings.
For example, tapping the ``\emph{Files}'' button (\autoref{fig:pomlink-table-selector} (b)) initiates a pop-up window, asking the user to select multiple documents or tables from the file system (\autoref{fig:pomlink-table-selector} (c)), with an optionally adjusted inclusion style, which are then linked to the agent's session.
Once files are linked, users can have a conversation with the agent (\autoref{fig:teaser} (2); \autoref{fig:pomlink-table-selector} (1)).
This interface enables both customized ``ask anything'' queries and the use of predefined task shortcuts like ``Translate'', ``Summarize'', ``Rewrite'', and ``Brainstorm''.
These shortcuts operate on the context provided by \emph{both} the linked files and the rich multi-modal chat history.
The \textbf{primary objectives} of this case study were twofold: first, to validate POML's effectiveness by integrating it into a functional mobile application prototype; and second, to assess how POML's features aided the development process, particularly in handling diverse data inputs and managing prompt complexity within this application.

PomLink's core agent functionality relies extensively on POML for structuring the prompts used to interact with LLMs.
This structure incorporated several key POML features, visible in the example prompt within the development environment shown in \autoref{fig:teaser} (b).
Specifically, we used:

\begin{itemize}
  \item An \texttt{<include>} component to modularly incorporate a common system prompt stored in a separate file, enhancing reusability and maintainability.
  \item The \texttt{<conversation>} component to present the conversational history between the user and the agent systematically, crucial for maintaining context in interactive sessions.
  \item Components like \texttt{<role>}, \texttt{<task>}, and \texttt{<output-format>} to provide a clear structure based on the RTF prompting framework \cite{rtfprompt}, organizing instructions for both the LLM and human developers.
  \item Various data components (\S~\ref{sec:poml_data_components}) to integrate content from the linked files selected by the user.
  \item The styling system (\S~\ref{sec:poml_styles}) to define formatting rules centrally and create distinct style profiles (e.g., compact vs. verbose), allowing adaptation to specific tasks or backend LLMs.
\end{itemize}

Integrating varied data sources into the prompt context is a key challenge in developing applications like PomLink.
POML addresses this by providing components to reference diverse data types structurally and the flexibility to specify their processing and presentation to the LLM.
For instance, the \texttt{<document>} component was used to embed content from user-selected text document files (e.g., PDF documents).
Attributes within this component allowed control over details like page selection or whether to preserve original formatting and embedded images.
Similarly, the \texttt{<table>} component ingested data from table files (e.g., Excel or CSV).
This component allowed specifying input formats and ensured consistent rendering of table data (e.g., as Markdown or CSV) within the prompt.
\autoref{fig:pomlink-table-selector} (2) displays the user-facing options derived directly from POML's \texttt{<table>} component capabilities for customizing data processing, such as toggling file name inclusion (\autoref{fig:pomlink-table-selector} (d)), incorporating column descriptions, selecting rendering formats, and choosing between full or partial table rendering.
Notably, the ``Auto'' table format option (\autoref{fig:pomlink-table-selector} (e)) leverages findings from our TableQA study (\S~\ref{sec:cs_tableqa}) to automatically select the empirically determined best format for the target LLM.
Representing other data types like voice memos (\texttt{<audio>}), web pages (\texttt{<webpage>}), images (\texttt{<img>}), or managing conversation history (\texttt{<conversation>}) was achieved straightforwardly using POML's unified framework for context data aggregation.

PomLink was implemented using the React Native framework, a popular choice for cross-platform mobile application development.
The development process for PomLink highlighted POML's value for rapid prototyping.
A functional prototype was completed by \emph{one of the POML development team in just 2 days}.
Around 90\% of this time was dedicated to iOS environment configuration, UI development, and simulator deployment, rather than core prompt engineering tasks.
The application's core logic required \emph{only 6 POML prompts}: 1 common system prompt, 4 task-specific prompts, and 1 prompt processing ``ask-anything'' queries, averaging a concise \emph{35 lines of code} each.
This efficiency stems from POML's ability to abstract complex data handling.
The built-in data components managed file parsing and formatting, eliminating the need for custom data processing code  typically required in traditional approaches.
This allowed the developer to focus primarily on the application's logic and user interface.
Furthermore, prompt styling was managed effectively through POML's internal styling system; presentation adjustments (e.g., captions, spacing, syntax) were made efficiently by modifying the central stylesheet JSON (\autoref{fig:teaser} (c)), without altering the main prompt logic.

The development workflow for PomLink was significantly accelerated by POML's toolkit (\S~\ref{sec:toolkit}), particularly the VSCode extension.
Within the PomLink project, the extension's live preview (\autoref{fig:teaser} (g)) provided immediate visual feedback on prompt rendering, including embedded document files (\autoref{fig:teaser} (h)), while integrated diagnostics (\autoref{fig:teaser} (e)) instantly flagged errors, streamlining debugging.
Hover documentation (\autoref{fig:teaser} (d)) and auto-completion (\autoref{fig:teaser} (f)) reduced the need to consult external references.
The interactive testing feature (\autoref{fig:interactive-testing}) allowed rapid iteration by enabling direct LLM calls and response viewing within the editor, considerably shortening the debug cycle for PomLink's prompts.
Integration into the PomLink application's React Native backend was achieved using the POML Node.js SDK (\autoref{fig:sdks} (a)), which handled loading POML files, injecting dynamic data, and rendering the final prompt string.
The text-based, modular nature of POML prompts also integrated seamlessly with Git for version control within the project (\autoref{fig:teaser} (a)).

In summary, the PomLink case study validated POML's effectiveness as a practical tool for real-world LLM application development.
It demonstrated how POML's features streamline development: its structural markup (\textbf{DG1}), its data components (\textbf{DG2}), styling system (\textbf{DG3}), and development toolkit (\textbf{DG4}) collectively enabled rapid prototyping and simplified the management of complex, multi-modal prompts.
The ability to create a versatile agent application prototype in just two days highlights POML's potential to accelerate innovation in developing sophisticated LLM-powered software by managing prompt structure and integration complexities effectively.

\begin{figure}[t]
  \centering
  \includegraphics[width=\linewidth]{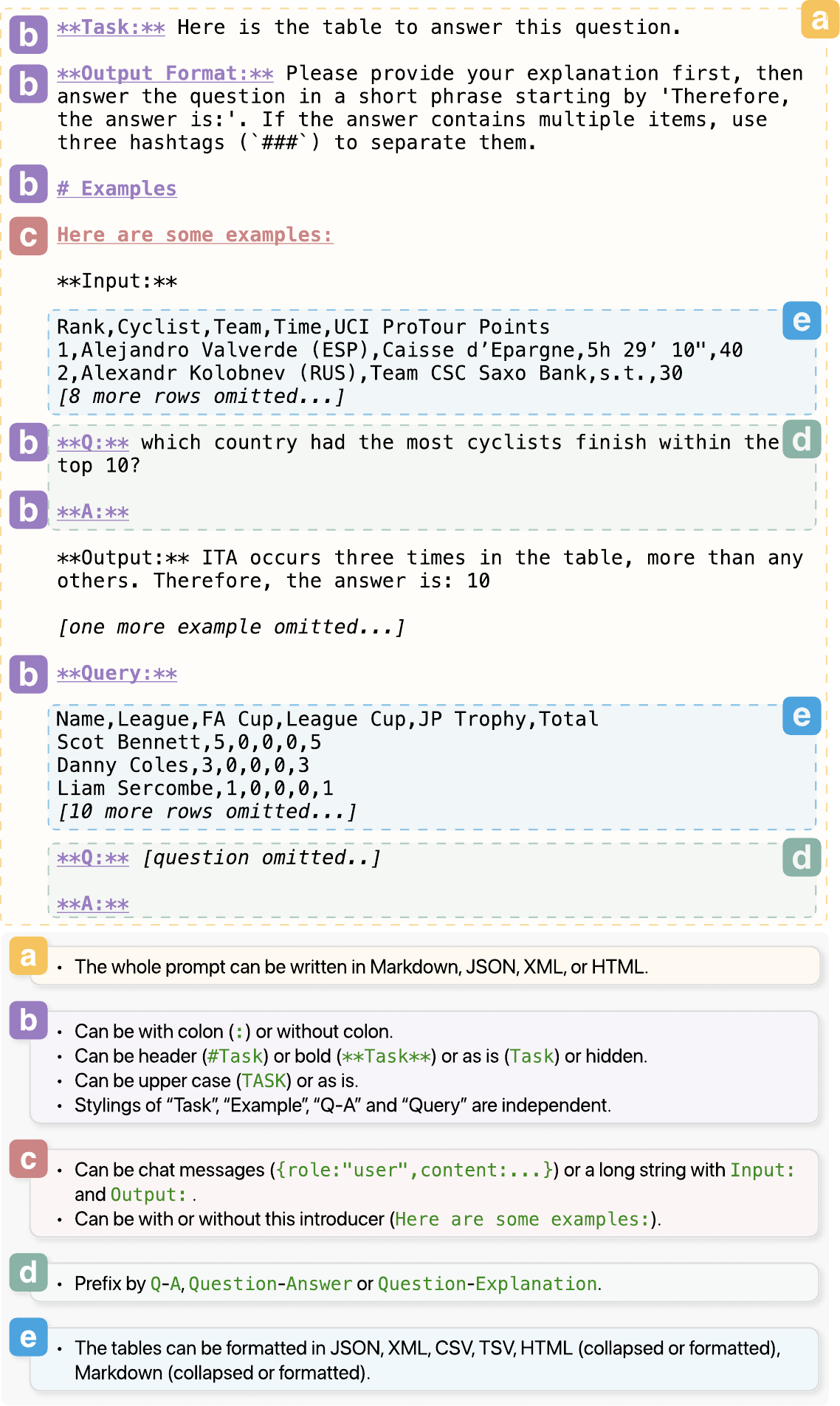}
  \caption{Visualization of the prompt styling search space explored in our TableQA experiment.
  The diagram illustrates the dimensions of prompt variation investigated, including overall syntax structure (a), section formatting options (b, d), example presentation styles (c), and table representation alternatives (e).
  Systematically combining these choices resulted in an extensive search space of 74k unique prompt styles.}
  \label{fig:search_space}
\end{figure}

\subsection{TableQA -- Deep Exploration of Prompt Styling}
\label{sec:cs_tableqa}

\begin{table*}[t]
  \caption{Optimal prompt styling configurations that yielded the highest accuracy for each LLM on the WikiTQ subset.
  Shows the specific combination of syntax and formatting choices (e.g., Overall Syntax (\autoref{fig:search_space} (a)), Example Body format (\autoref{fig:search_space} (c)), Table Syntax (\autoref{fig:search_space} (e))) for the best-performing style among 100 sampled styles.}
  \label{tab:model_perf_summary}
  \resizebox{\textwidth}{!}{\begin{tabular}{l|cccccccc} 
\hline
\textbf{Model} & \textbf{Overall Syntax} & \textbf{Table Syntax} & \textbf{Instruction Header} & \textbf{Example Caption} & \textbf{Example Body} & \textbf{QA Caption} & \textbf{QA Body} & \textbf{User Input Caption} \\ \hline
Claude 3 Haiku & XML & HTML (Ugly) & -- & -- & -- & -- & -- & -- \\
DeepSeek V3 & Markdown & HTML & Plain-Upper & Hidden & Hidden & Bold-Colon & Question-Answer & Hidden \\
Gemini 2.0 Flash & HTML & HTML & Header-Upper & Hidden & Introducer & Bold-Colon & Q-A & Header-Upper \\
GPT-3.5 Turbo & Markdown & TSV & Plain-Upper-Colon & Header & Hidden & Hidden & Hidden & Hidden \\
GPT-4o Mini & Markdown & Markdown & Plain-Upper & Header & Hidden & Bold-Colon & Question-Answer & Plain-Upper \\
LLaMA 3 70B & Markdown & Markdown & Header-Colon & Header-Colon & Chat w. Introducer & Header-Colon & Question-Explanation & Header-Colon \\
Mistral AI 8x7B & Markdown & XML & Bold-Upper-Colon & Bold-Upper-Colon & Chat & Bold-Colon & Question-Answer & Bold-Upper-Colon \\
Phi-3 Medium & JSON & CSV & -- & -- & -- & -- & -- & -- \\
\hline
\end{tabular}
}
\end{table*}

\begin{table}[t]
  \caption{Impact of prompt styling variations on LLM accuracy for TableQA (WikiTQ subset). Shows minimum/maximum accuracy across 100 sampled styles, performance difference (Diff), relative improvement ((Max-Min)/Min), and style ranking stability (Self-correlation: mean Spearman correlation of style rankings across 1000 random data splits, see \autoref{fig:correlation_heatmap}).}
  \label{tab:model_perf_best}
  \resizebox{\linewidth}{!}{\begin{tabular}{l|ccc||c} 
\hline
\textbf{Model} & \textbf{Acc Min} & \textbf{Acc Max} & \textbf{Diff} & \textbf{Self-corr.} \\ \hline
Claude 3 Haiku & 0.138 & 0.555 & 0.417 (\textcolor{blue}{+303\%}) & 0.866  \\
DeepSeek V3 & 0.682 & 0.823 & 0.141 (\textcolor{blue}{+21\%}) & 0.348  \\
Gemini 2.0 Flash & 0.717 & 0.830 & 0.113 (\textcolor{blue}{+16\%}) & 0.314  \\
GPT-3.5 Turbo & 0.060 & 0.618 & 0.558 (\textcolor{blue}{+929\%}) & 0.767  \\
GPT-4o Mini & 0.622 & 0.753 & 0.131 (\textcolor{blue}{+21\%}) & 0.473  \\
LLaMA 3 70B & 0.152 & 0.657 & 0.505 (\textcolor{blue}{+333\%}) & 0.606  \\
Mistral AI 8x7B & 0.177 & 0.481 & 0.304 (\textcolor{blue}{+172\%}) & 0.849  \\
Phi-3 Medium & 0.007 & 0.322 & 0.314 (\textcolor{blue}{+4450\%}) & 0.931  \\
\hline
\end{tabular}
}
\end{table}

Prior research indicates that the specific structure and format used to present information within a prompt can significantly influence LLM performance \cite{sclarquantifying, salinas2024butterfly, voronov2024mind}.
However, systematically exploring this styling aspect is challenging due to the combinatorial complexity of generating prompt variants.
This case study investigates the impact of prompt styling on LLM performance for the task of table-based question answering (TableQA).
Our \textbf{primary goals} were twofold: first, to quantify the sensitivity of different LLMs to stylistic variations in prompts, and second, to demonstrate how POML's features, particularly its styling system, facilitate systematic management and exploration of these prompt styles.

To conduct this investigation, we leveraged POML's separation of content (markup) and presentation (stylesheets).
We authored a single base POML prompt defining the TableQA task structure, including intention components, few-shot examples, and placeholders for the question and table data (provided in Appendix~\ref{sec:tableqa_prompt}).
We then defined a set of stylistic variations within POML stylesheets (example also available in Appendix~\ref{sec:tableqa_prompt}).
These stylesheets controlled elements like the overall prompt syntax (e.g., Markdown, JSON, XML), the rendering format of embedded table data (e.g., CSV, Markdown, HTML), the presentation style of instructions and examples (e.g., caption styles, visibility, chat formatting), and the structure of the question-answer section, as visualized in \autoref{fig:search_space}.

From this extensive space, we randomly sampled 100 styles without replacement for evaluation.
These 100 distinct prompt styles were tested on 283 samples (10\% of the data) from the WikiTableQuestions (WikiTQ) validation dataset \cite{pasupat2015compositional}.
Accuracy was measured by comparing the LLM's generated answer and the ground truth answer with evaluation tools provided by WikiTQ.
The evaluation included 8 low-cost LLMs (priced under \$0.3 per million input tokens as of February 2025): Claude-3-Haiku \cite{claude3}, DeepSeek-V3 \cite{deepseekai2024deepseekv3}, Gemini-2.0-flash \cite{gemini2}, GPT-3.5-turbo-0125 \cite{gpt35turbo}, GPT-4o-mini \cite{gpt4o}, LLaMA3-70B \cite{grattafiori2024llama3herdmodels}, Mistral-AI-8x7b \cite{jiang2024mixtralexperts}, and Phi3-medium \cite{abdin2024phi3technicalreporthighly}.

\paragraph{Results}
The experimental results confirmed a significant dependency between prompt styling and TableQA accuracy, as detailed in \autoref{tab:model_perf_best}.
The degree of sensitivity to styling, however, varied considerably across models.
Some models were extremely sensitive: GPT-3.5-Turbo's accuracy ranged from 6\% to 61.8\%, a relative improvement of 929\%, while Phi-3 Medium improved by 4450\% (from 0.7\% to 32.2\% accuracy) between its worst and best styles.
Other models, such as DeepSeek V3 and Gemini 2.0 Flash, were less sensitive, with performance varying by 21\% and 16\% respectively, highlighting model-dependent sensitivity.
The self-correlation scores in \autoref{tab:model_perf_best} (especially 0.931 for Phi-3 Medium, 0.866 for Claude 3 Haiku) suggest that the relative performance ranking of different styles is stable for many models across different data subsets.
Our further analysis revealed that the optimal prompt style is model-specific.
\autoref{tab:model_perf_summary} details the combination of styling features (e.g., overall syntax, table syntax, example formatting) that yielded the best performance for each model.
This model-specific optimal styling highlights the challenge of prompt portability and the need for adaptable styling mechanisms.
Overall, this TableQA case study underscores the critical role of prompt styling in achieving optimal LLM performance, which POML addresses.

\begin{table}[t]
  \caption{Table format preferences across LLMs for the TableQA task. For each model, the table shows the top two preferred table syntax formats (e.g., CSV, Markdown, HTML) based on the highest average exact match accuracy achieved by prompt styles utilizing that specific format within our sample of 100 styles. The corresponding mean accuracy for each format is shown in parentheses.}
  \label{tab:table_format_pref}
  \resizebox{\linewidth}{!}{\begin{tabular}{lll}
\toprule
\textbf{Model} & \textbf{Best Format} & \textbf{Second Best Format} \\
\midrule
Claude 3 Haiku & CSV (\textcolor{blue}{0.449}) & Markdown (\textcolor{blue}{0.424}) \\
DeepSeek V3 & HTML (\textcolor{blue}{0.801}) & JSON (\textcolor{blue}{0.800}) \\
Gemini 2.0 Flash & XML (\textcolor{blue}{0.791}) & HTML (\textcolor{blue}{0.788}) \\
GPT-3.5 Turbo & Markdown (Collapse) (\textcolor{blue}{0.524}) & CSV (\textcolor{blue}{0.510}) \\
GPT-4o Mini & JSON (\textcolor{blue}{0.702}) & Markdown (\textcolor{blue}{0.694}) \\
LLaMA 3 70B & Markdown (\textcolor{blue}{0.601}) & HTML (\textcolor{blue}{0.601}) \\
Mistral AI 8x7B & XML (\textcolor{blue}{0.365}) & HTML (Ugly) (\textcolor{blue}{0.348}) \\
Phi-3 Medium & CSV (\textcolor{blue}{0.169}) & JSON (\textcolor{blue}{0.116}) \\
\bottomrule
\end{tabular}}
\end{table}

A critical styling dimension for TableQA is the format used to represent the table data itself.
Our results, summarized in \autoref{tab:table_format_pref}, reveal diversity in the optimal table formats across different models, consistent with previous findings \cite{zhang2024benchmarking, he2025tableloralowrankadaptationtable}.
While some models performed best with simple formats like CSV, others preferred structured representations like HTML, XML, JSON, or Markdown variants.
This variation emphasizes the importance of tailoring table representations to the target LLM for optimal performance.
These findings directly informed the development of our PomLink application (\S~\ref{sec:cs_pomlink}).
The ``Auto'' table format option within PomLink uses these results, automatically selecting the table syntax identified as optimal for the user's chosen backend LLM based on this study.

On the other hand, the experiment demonstrates the value of POML's design, especially the separation of content (markup) from presentation (stylesheets) (\textbf{DG3}).
This study required only one single, concise base POML prompt (30 lines of code, see Appendix~\ref{sec:tableqa_prompt}).
By programmatically combining stylesheet variations, we generated a combinatorial search space encompassing 73,926 unique prompt style configurations derived from the single base POML file.
Using the POML Python SDK (\S~\ref{sec:sdks}), we loaded the base ``\texttt{.poml}'' file and rendered final prompts by combining it with the generated stylesheets and the specific table/question data for each test case.
By enabling systematic variation and optimization of styling parameters independently of the core prompt logic, POML provides a mechanism for adapting prompts to different LLM characteristics and maximizing performance through experimentation, reducing the engineering effort compared to manually managing numerous prompt variations.

\section{User Study}
\label{sec:user_study}

We conducted a formal user study to evaluate the usability, effectiveness, and developer experience of POML and its toolkit in practical scenarios.
The study aimed to assess POML's utility for representative prompt engineering tasks and gather qualitative feedback to identify its strengths and limitations.

\paragraph{Participants}
We recruited seven participants with diverse technical backgrounds, including software engineers, researchers, and students.
Participants had no prior exposure to POML or its use cases before the study.
Participant backgrounds, self-reported prompt engineering experience levels, and prior use of LLMs in application development are detailed in \autoref{tab:user_study_completion}.
This diversity was sought to gather feedback reflecting different potential user experiences and needs, especially regarding prior experience developing LLM-related applications.

\paragraph{Tasks}
We designed five distinct tasks of increasing complexity to probe different facets of POML's capabilities.

\begin{itemize}
\item \textbf{Task 1 (T1)} involved rewriting an existing plain text prompt into POML and subsequently restyling its output presentation to YAML format using POML's styling features.
\item \textbf{Task 2 (T2)} focused on utilizing POML's document handling features (\texttt{<document>}) to process TODO items embedded within a Microsoft Word document, showcasing its ability to integrate and present document content effectively.
\item \textbf{Task 3 (T3)} required participants to analyze stock market data (alongside its visualization) provided in an Excel spreadsheet (\texttt{<table>}) and prompt an LLM for investing recommendations, testing table integration and image presentation.
\item \textbf{Task 4 (T4)} explored meta-prompting concepts by asking participants to use an LLM (initially GPT-4o) to generate POML code based on a list of POML examples; they were then asked to adapt the prompt to target a smaller code-completion model, CodeGemma-2B \cite{codegemmateam2024codegemmaopencodemodels}, evaluating POML's role in managing prompts for different models and the perceived difficulty of this adaptation.
\item \textbf{Task 5 (T5)}, the most complex task, involved translating content from two subtitle files (in TSV format, one from a 22-minute animation, the other from a 45-minute news report) while preserving specific formatting conventions and timestamps, testing POML's ability to read, process, and present tabular data subsets in specific formats under constraints.
\end{itemize}

The code and data associated with these tasks are available in the supplementary material.

\paragraph{Procedure}
Each participant engaged in a single session lasting approximately 90 minutes.
To manage session time effectively, participants were assigned a randomly selected subset of the five tasks.
Participants used their own laptops, requiring only VSCode, without additional runtime environment requirements.
Sessions began with participants watching a 7-minute introductory video tutorial explaining POML's core concepts and usage.
Subsequently, they were provided with two quickstart code examples and a concise Markdown manual detailing POML syntax and features.
They were provided with a VSCode workspace containing the task descriptions, data, and preconfigured access to the GPT-4o and CodeGemma-2B LLMs.
Participants were instructed to use a think-aloud protocol, verbalizing their thought processes, challenges, and insights while completing the assigned tasks within the provided VSCode environment equipped with the POML language extension.
Following the task completion phase, participants engaged in a semi-structured verbal interview guided by a predefined list of 11 key questions.
These questions explored their task experience (completion, time, difficulties), prior prompt usage, perceived usefulness of POML (scenarios, valuable/difficult features, learnability), potential workflow integration, feedback on the VSCode tooling, suggestions for improvement, and any encountered bugs (the full list is provided in Appendix~\ref{sec:interview_questions}).

All user study sessions were screen and audio recorded for subsequent detailed analysis.
Comprehensive telemetry data was automatically logged, including the time spent on each task, interactions with Language Server Protocol (LSP) features (such as hover help activations, code completion triggers and acceptances), and the complete version history of the POML code written by each participant.
Facilitators maintained a minimal intervention policy, offering assistance only when a participant encountered a significant impediment preventing further progress.
The evaluation involved a mixed-methods approach, combining qualitative analysis of the verbal feedback and facilitator observations with quantitative analysis of task completion metrics and telemetry data.

\subsection{Task Completion and Component Usages}
\label{sec:results_performance}

\begin{table*}
  \centering
  \caption{User study results showing task completion and tool interaction metrics. \textbf{Left Columns:} Participant backgrounds. \textbf{Middle Columns:} Task completion status (\mycheckmark = completed, \mypartial = partially completed, \mydash = not assigned) and approximate completion times. \textbf{Right Columns:} Interaction metrics (hover help usage, code completion attempts, items per suggestion, suggestion acceptance rate). Task complexity increased from T1 (prompt rewriting) to T5 (subtitle translation), with T5 requiring more time. Participants showed varied engagement with IDE assistance features.}
  \label{tab:user_study_completion}
  \resizebox{\textwidth}{!}{\begin{tabular}{c|cc||ccccc||cccc}
\hline
\multirow{2}{*}{\textbf{Participant}} & \multicolumn{2}{c||}{\textbf{Prompt Experience}} & \multicolumn{5}{c||}{\textbf{Task Completion}} & \multicolumn{4}{c}{\textbf{LSP Metrics}} \\
\cline{2-3} \cline{4-8} \cline{9-12}
& Rate & LLM-App-related & T1 & T2 & T3 & T4 & T5 & Hover & Completion & Items/Comp & Accept (\%) \\
\hline
P1 & Novice & No & \mydash & \mycheckmark ~ 10min & \mycheckmark ~ 10min & \mydash & \mycheckmark ~ 1h & 109 & 908 & 2.38 & 21.21 \\
P2 & Advanced & Yes & \mycheckmark ~ 20min & \mydash & \mycheckmark ~ 15min & \mypartial ~~ 20min & \mydash & 68 & 286 & 2.18 & 49.37 \\
P3 & Intermediate & No & \mydash & \mycheckmark ~ 10min & \mydash & \mydash & \mycheckmark ~ 1h & 35 & 402 & 3.00 & 57.14 \\
P4 & Intermediate & No & \mydash & \mycheckmark ~ 10min & \mycheckmark ~ 10min & \mypartial ~~ 10min & \mydash & 24 & 226 & 2.50 & 50.00 \\
P5 & Novice & No & \mycheckmark ~ 10min & \mycheckmark ~ 5min & \mydash & \mydash & \mypartial ~~ 40min & 21 & 544 & 2.55 & 27.27 \\
P6 & Advanced & Yes & \mydash & \mycheckmark ~ 10min & \mycheckmark ~ 15min & \mycheckmark ~ 20min & \mydash & 120 & 1282 & 2.11 & 23.66 \\
P7 & Intermediate & Yes & \mydash & \mycheckmark ~ 10min & \mycheckmark ~ 15min & \mycheckmark ~ 15min & \mypartial ~~ 20min & 22 & 1734 & 2.16 & 50.82 \\
\hline
\textbf{Average} & -- & -- & 15min & 9min & 13min & 16min & 45min & 57.0 & 768.9 & 2.41 & 39.92 \\
\hline
\end{tabular}
}
\end{table*}

\begin{figure}
  \centering
  \includegraphics[width=\linewidth]{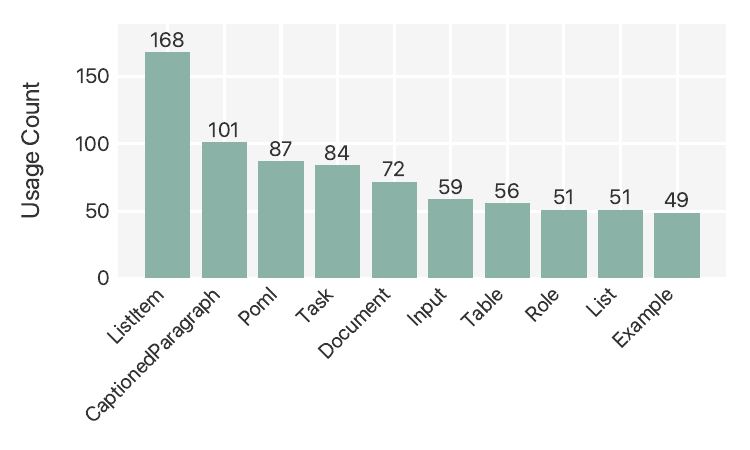}
  \vspace{-5mm}
  \caption{Frequency of POML component usage across user study sessions. The chart shows the total count of each component type used by participants. List items (\texttt{<item>}) were most frequently used, followed by captioned paragraphs (\texttt{<cp>}), and the root \texttt{<poml>} tag. Data components (\texttt{<document>}, \texttt{<table>}) and intention components (\texttt{<task>}, \texttt{<role>}) were also commonly used, showing engagement with POML's data and intention components.}
  \label{fig:user_study_components}
\end{figure}

Participants demonstrated high overall task completion rates across the assigned activities.
\autoref{tab:user_study_completion} details the completion status for each participant across the five tasks.
Average task completion times varied considerably based on task complexity and participant differences, as also shown in \autoref{tab:user_study_completion}.
Simpler tasks, such as T1 (Rewriting/Restyling) and T2 (Document TODOs), were typically completed within approximately 10--15 minutes by assigned participants, suggesting effective basic usability.

Tasks evaluating fundamental POML capabilities were generally completed successfully.
Participants generally completed T1 and T2 relatively quickly.
T2 specifically demonstrated the perceived value of POML's document handling features, although one participant (P7) noted occasional performance lag with larger documents.
T3 (Stock Analysis) was also successfully completed by most assigned participants, effectively demonstrating POML's integration and presentation capabilities for tabular data.
Some participants, however, encountered minor difficulties with the specific syntax for selecting table data subsets in T3.
T4 (Meta-Prompting) presented greater challenges.
While most participants successfully generated initial POML code using GPT-4o, adapting the prompt for the smaller CodeGemma-2B model proved difficult within the session's time constraints, reflecting inherent challenges in model-specific prompt adaptation.
T5 (Subtitle Translation) represented the most complex task, exhibiting the lowest completion rate and the longest average completion times among assigned participants, as detailed in \autoref{tab:user_study_completion}.
Common difficulties included managing precise TSV formatting, handling potential token limits with the longer input file, and preventing undesired line merging or timestamp inaccuracies.
This task effectively probed the limits of managing complex format preservation under constraints using POML.

Analysis of component usage frequency, illustrated in \autoref{fig:user_study_components}, reveals participants' engagement with POML's core structuring elements.
Basic structural components like list items (\texttt{<item>}) and captioned paragraphs (\texttt{<cp>}) were the most frequently employed after the root \texttt{<poml>} tag.
Intention components (e.g., \texttt{<task>}, \texttt{<input>}, \texttt{<role>}, \texttt{<example>}) and data integration components (\texttt{<document>}, \texttt{<table>}) also saw significant usage.
This pattern indicates that participants actively used POML's features for structured prompting and data handling throughout the study tasks.

Interaction metrics, presented in the rightmost columns of \autoref{tab:user_study_completion}, indicate diverse patterns of reliance on the provided IDE tooling assistance.
We observe that hover help usage varied widely among individuals, while code completion features were frequently triggered by most participants.
However, the acceptance rates for completion suggestions (ranging from 21\% to 57\%) suggest mixed perceived utility or accuracy of the LSP-provided suggestions.
Observations during the sessions confirmed differing programming habits: some participants (P3, P4, P7) used auto-completion extensively, while others (P6) frequently used the hover feature to explore documentation, and some relied more on manual typing or consulting the provided documentation.

\subsection{User Experience and Qualitative Feedback}
\label{sec:results_experience}

Participants' experiences with POML were largely positive, shaped by their technical backgrounds and task complexity.
Feedback consistently highlighted POML's utility in structuring prompts and integrating data, while also pinpointing areas for potential refinement (\S~\ref{sec:discussion}).

\paragraph{Feedback on Core Language Features}
POML's capability for integrating diverse data types emerged as a frequently praised strength.
Participants consistently valued the built-in components for handling various file types (\texttt{<document>}, \texttt{<table>}, etc.), emphasizing the reduction in manual data preparation effort compared to traditional methods.
For example, P7 noted the convenience of reading ``various different files... PDFs, Word documents, and Excel files,'' while P3 found features like ``selected records'' and ``selected columns ... quite helpful.''
P4 appreciated how POML simplified previously ``troublesome'' tasks like ``reading documents... reading images,'' enabling quick data loading via simple commands.
This positive reception aligns with the observed usage of data components in tasks T2, T3, and T5, as shown in \autoref{fig:user_study_components}.

The styling system was also well-received.
Participants recognized the value of format control (P1 found the format changing feature ``quite cool,'' a sentiment echoed by P7 regarding the value of format control).
They also praised its suitability for structured tasks like format conversion.
P5 described POML as ``suitable for structured work... especially for file format conversion.''.
While tasks within the study can often be resolved with inline attributes only (e.g., in T1, T3, T5), P2 successfully used an external \texttt{<stylesheet>} in T1, demonstrating the feasibility of the approach.
However, deeper engagement with complex stylesheets was limited by the study's scope, leaving the full potential for systematic format management largely recognized conceptually.

\paragraph{Development Experience}
The integrated development toolkit received predominantly positive feedback for enhancing the prompt authoring workflow.
The live preview feature was universally praised for providing immediate visual feedback on the rendered prompt structure, thereby reducing cognitive load and improving clarity (P1, P4).
The integrated testing feature, particularly its real-time streaming of LLM outputs, was also valued, with P6 finding the streaming capability ``surprisingly impressive.''
Participants frequently adopted an iterative edit-preview-test cycle within the IDE, consistent with common prompt engineering practices \cite{zamfirescu2023johnny}.
Usage patterns for IDE assistance features varied, as discussed in \S~\ref{sec:results_performance}.
Hover help was frequently used by some participants.
For instance, P6 mentioned ``frequently using the hover feature to explore available documentation''.
Auto-completion saw extensive use by others (e.g., P3, P4, P7), indicating active engagement with these assistance tools despite differing individual habits.

\paragraph{Learning Curve}
Participants' prior technical experience, particularly familiarity with HTML/XML markup, significantly influenced their perceived learning curve.
Those with relevant backgrounds (P2 found it ``elegant like React''; P7 was ``familiar with markup languages'') generally found the syntax intuitive.
Conversely, participants without such experience (P1 stated ``I don't know HTML very well ... don't know attributes need to be quoted'') reported a steeper initial learning curve and relied more heavily on the live preview and provided documentation.
P6 initially found the number of components somewhat ``overwhelming.''
The provided documentation (video, examples, manual) served as crucial initial learning resources, frequently consulted by participants (especially P1, P3, P6).
Despite varying adaptation speeds, all participants appeared to gain comfort with basic POML usage within 10--15 minutes of starting their first task, aided by the IDE's live preview and diagnostics.

\paragraph{Workflow Integration}
Participants described diverse existing strategies for managing prompts, ranging from informal methods like using note-taking applications (P1 stated ``I currently manage my prompts through OneNote'') or simple text files (P5), to embedding prompts directly within application code (P2 mentioned ``If I write in Python, I often write prompts messily [in Python string]'').
Despite these varied habits, a majority (5 of 7) expressed potential for integrating POML into their workflows, particularly for two types of use cases.
First, for managing frequently used or complex \emph{personal prompts}, participants valued the organization POML provides (P1 explained ``if I put them in a [code] repo, I can find them,'' contrasting with the disorganization of OneNote).
Second, for \emph{developing LLM applications}, participants with LLM-App development experience saw significant value in treating POML files like source code.
This approach facilitates standard version control practices (P2 and P6 explicitly mentioned ``Git integration'') and improves collaboration through shared, standardized prompt definitions (P6 reflected that ``it might be more conducive to my subsequent prompt management'').
Integration with Python, though not reflected in the given tasks, was frequently cited as critical for adoption in application development contexts (P1, P5, P6), as well as programmatic prompt generation and incorporation into existing AI/ML pipelines.

\paragraph{Envisioned Values}
Beyond the study tasks, participants envisioned several practical applications where POML's structure and data handling would be beneficial.
P1 suggested using POML to manage context for writing tasks by embedding related research papers as documents: ``let it read several related works first... then let it follow some instructions to write [my article].''
P3 similarly saw potential for academic writing support.
P2 highlighted the convenience for table processing, stating that without POML, it ``would require writing a lot of code and formatting... but with POML it would be convenient,'' envisioning its use for ``running benchmarks.''

\paragraph{Unexpected Usage Patterns and Desires}
Observations revealed potential use cases beyond POML's primary design focus.
Some participants (P1, P3, P6) appeared to use data components (e.g., \texttt{<document>}) partly as a convenient way to preview file contents within the IDE, suggesting secondary value as an integrated data viewer.
P3 appreciated how POML could ``import a file directly, then help me read it out, help me display it,''
Occasional comments also hinted at a desire among some advanced users (P6, P7) for lower-level customization options regarding component rendering.
Interest was also expressed in exploring agent-like capabilities by writing multiple distinct prompts within a single \texttt{.poml} file, with P5 suggesting ``potential for extending the framework towards more complex task orchestration.''

\paragraph{Usage Overhead}
The frequent use of intention and structural components (\autoref{fig:user_study_components}) suggests that most participants appreciated the explicit structure POML imposes for managing prompt complexity in such scenarios.
However, a trade-off regarding usage overhead was noted by several participants (P1, P4, P6).
For simple, one-off prompts, the explicit component structure could introduce typing or cognitive overhead compared to plain text, described by P1 and echoed by P5 as potentially ``using a sledgehammer to crack a nut.''
While POML's design attempts to minimize this for simple cases (e.g., the root \texttt{<poml>} tag is optional, as discussed in \S~\ref{sec:poml_structure}), the need to create a file and the mental burden caused by \texttt{.poml} suffix still represents a higher initial cost than typing directly into a chat interface.
This feedback suggests POML's value proposition is strongest for complex, data-intensive prompts where structure, lifecycle maintainability, and reusability are paramount, while it may be overkill for simple, transient plain text prompts.

\section{Discussion}
\label{sec:discussion}

The case studies and user study offered practical insights into POML's application and adoption.
Although POML's strengths were confirmed, the evaluation also highlighted areas for refinement based on user feedback and task challenges.
These findings inform this discussion on improvements, limitations, and future directions.

\subsection{Suggestions for Improvement}
\label{sec:discussion_suggestions}

\paragraph{Usability and Developer Experience Improvements}
Participants suggested several usability enhancements to improve the learning curve and development workflow.
A recurring request was for more comprehensive documentation, including searchable references and practical examples covering advanced styling and templating (P2 asked for ``documentation that's easy to look up'').
Clearer, more actionable error messages were desired to expedite debugging, as current messages were sometimes found opaque (P1 and P3 stated ``Error messages are hard to understand'').
Language or tooling simplifications, such as sensible default attributes (P2) or clearer distinctions between similar components (P6), were also proposed.
Performance optimization, particularly for large document handling (P7) and code assistance responsiveness, remains crucial.
Refining auto-completion accuracy (P7) and mitigating conflicts with tools like GitHub Copilot (P2) were noted as important for a smoother experience.

\paragraph{Feature Requests}
User requests focused on enhancing workflow integration and expanding POML's functionality.
Participants desired options to directly save LLM test outputs to files (P6), bypassing manual copying.
Improved mechanisms for managing multi-turn conversations within the POML tool were commonly sought by almost all participants.
Greater user control over LLM selection and API key management during testing was also requested (P3, P5).
Support for additional document formats like LaTeX, particularly for academic use cases (P3), was suggested.
Some novice users indicated a desire for integrated guidance on prompt best practices or model-specific formatting (P1), reflecting known challenges in prompt engineering interfaces \cite{zamfirescu2023johnny}.

\subsection{Limitations}
\label{sec:limitations}

\paragraph{Accessibility}
The current POML VSCode extension (especially the live preview) presents significant accessibility barriers.
It lacks optimization for users relying on assistive technologies like screen readers or requiring high-contrast modes.
Key areas needing improvement include keyboard navigability, text equivalents for UI elements, sufficient color contrast, and font size scaling support.
Addressing these gaps based on established guidelines \cite{w3caccessibility} is essential to ensure broader usability.

\paragraph{User Study}
Our user study findings (\S~\ref{sec:user_study}) are subject to methodological limitations.
The 90-minute sessions restricted deep exploration of advanced features.
The small participant sample ($N=7$), though diverse, consisted mainly of technical users, potentially limiting generalizability.
The controlled lab setting might not fully reflect real-world project complexities.
Furthermore, most participants did not deeply engage with complex stylesheets during the tasks.
To help mitigate these limitations, we are actively working to open-source POML on GitHub and release the extension on the VSCode Extension Marketplace, which will allow us to gather feedback from a broader and more diverse audience using POML in real-world contexts.

\subsection{Future Work}
\label{sec:future_work}

Based on user feedback, case study insights, and identified limitations, future work will prioritize enhancing POML's usability, features, and ecosystem.
We will keep improving the developer experience through better accessibility support, augmented documentation, refined error diagnostics, performance optimizations, and enhanced SDK usability.
Key feature developments include more sophisticated multi-turn conversation management, flexible output handling (e.g., direct file saving), and an expanded template library, potentially incorporating model-specific styling insights (\S~\ref{sec:cs_tableqa}).
The structured nature of POML also makes it suitable for exploring automated prompt engineering techniques \cite{liu2025promptcontentenhancingllm, zhoularge, li2024guiding}, which is another promising direction for future research.

Beyond core enhancements, we aim to promote POML adoption and demonstrate its value across various domains.
We envision POML as a general-purpose prompt markup language.
Potential applications include serving as a meta-prompt language in agent systems \cite{OpenManus}, a structured approach to configure applications like Cursor \cite{cursormdc}, or a facilitator for industry-level LLM response evaluation workflows as seen in tools like the Azure AI SDK \cite{azureevaluationsdk}.
Further investigation into its utility in complex RAG pipelines, educational tools, domain-specific component libraries, and collaborative prompt engineering environments is also warranted.

\section{Conclusion}
\label{sec:conclusion}

This paper presented POML, a novel Prompt Markup Language designed to address critical challenges in contemporary prompt engineering.
POML introduces a component-based structure for enhanced clarity and maintainability, specialized components for effective integration of diverse data types, and a decoupled styling system to systematically manage LLM format sensitivity.
Complementing the language, an integrated development toolkit, featuring IDE support and multi-language SDKs, aims to streamline prompt authoring, testing, and management.
The effectiveness and practical utility of POML were empirically validated through two distinct case studies alongside a formal user study assessing usability and developer experience.
Collectively, these contributions establish POML as a structured, maintainable, and versatile paradigm that addresses prevalent prompt engineering difficulties, thereby enhancing developer workflows, prompt reusability, and collaborative efforts in building sophisticated LLM applications.

\bibliographystyle{ACM-Reference-Format}
\bibliography{main}

\appendix
\section{POML vs. Other Prompt Markup Languages}
\label{sec:related_work_compare}

\newcolumntype{B}{>{\hsize=1.6\hsize}X}
\newcolumntype{w}{>{\hsize=1.2\hsize}X}
\newcolumntype{s}{>{\hsize=0.8\hsize}X}
\newcolumntype{x}{>{\hsize=0.6\hsize}X}

\begin{table*}[t]
    \caption{Comparison of POML with other Prompt Markup Languages} 
    \label{tab:prompt_tool_comparison} 

    \begin{tabularx}{\linewidth}{%
         X 
         x  X  x  B  w  s  w  
    }
    \hline
    \textbf{Markup Language} & \textbf{Markup Syntax} & \textbf{Runtime Env.} & \textbf{Selling Point} & \textbf{Key Components} & \textbf{Data Support} & \textbf{Styling} & \textbf{Dev Tooling} \\
    \hline
    \textbf{AI.JSX} \cite{aijsx} & JSX/TSX & Node.js & Agents \& Workflows & High-level components (Chat completion, Tool use, Q\&A) & Relies on external libraries & --- & NextJS support, LangChainJS integration \\
    \hline
    \textbf{GenSX} \cite{gensx} & JSX/TSX & Node.js & Agents \& Workflows & High-level workflow/agent components & Text & --- & TypeScript tooling \\
    \hline
    \textbf{VSCode Prompt TSX} \cite{vscodeprompttsx} & TSX & Node.js (VS Code Ext) & Prompt (VS Code) & Role components (User/Asst), Tool (\texttt{<ToolResult>}), File (\texttt{<FileLink>}) integration & Chat history, VS Code API data (files), Tool results & Priority system (manages length) & VS Code Ext dev env (TSX/TS support) \\
    \hline
    \textbf{Priompt} \cite{priompt} & JSX/TSX & Node.js & Writing Prompt & Message \& Structural components (\texttt{<scope>}, \texttt{<first>}, etc.); Priority system (\texttt{p/prel}) & Image support, JSON data & Priority system (manages length) & Preview WebUI \\
    \hline
    \textbf{Prxmpt} \cite{prxmpt} & JSX & Node.js & Writing Prompt & Utility elements (\texttt{<kv>}, etc.), Priority (\texttt{<priority>}), Space control, Data serialization (\texttt{<json>}, \texttt{<yaml>}) & Primitives (\texttt{<num>}), Dates (\texttt{<datetime>}), Objects (\texttt{<json>}, \texttt{<yaml>}) & Basic text format (\texttt{<b>}, \texttt{<h1>}, etc.); Space control & JSX/TS support, Standard Node.js tooling \\
    \hline
    \textbf{MDXPrompt} \cite{mdxprompt} & MDX (Markdown + JSX) & Node.js & Writing Prompt & Mix structured/unstructured text, Default components (\texttt{<Prompt>}, \texttt{<ChatHistory>}) & Chat history, JSON data & --- & MDX/React ecosystem \\
    \hline
    \textbf{ChatML (OpenAI)} \cite{chatml} & Custom format (Roles) & Language-Agnostic & Writing Prompt & Role-based messages (system, user, assistant, tool) & Text, Images & Customizable chat templates & OpenAI SDKs \\
    \hline
    \textbf{PromptML} \cite{promptml} & Custom DSL & Language-Agnostic & Writing Prompt & Intention components (\texttt{@context}, \texttt{@objective}, \texttt{@instructions}, etc.) & Text & --- & Python Parser \\
    \hline
    \textbf{SAMMO} \cite{schnabel2024symbolicpromptprogramsearch} & Python API + Markdown & Python & Writing Prompt & Programmatic manipulation, Structure-aware optimization algos & Relies on Python data libs & --- & Python dev env, LLM Runner integrations \\
    \hline
    \textbf{POML} & HTML-like DSL & \textbf{Language-agnostic} & Writing Prompt & \textbf{Versatile (basic structure + intention + data)} & \textbf{Comprehensive (tables, docs, webpages)} & \textbf{CSS-like styling system} & \textbf{Full VS Code IDE extension} \\
    \hline
\end{tabularx}
\end{table*}

There have been various markup languages and frameworks, each addressing specific needs, as summarized in \autoref{tab:prompt_tool_comparison}.

JSX/TSX-based tools like AI.JSX \cite{aijsx} and GenSX \cite{gensx} primarily target agent and workflow orchestration using high-level components within the Node.js ecosystem.
Others, such as VSCode Prompt TSX \cite{vscodeprompttsx}, Priompt \cite{priompt}, Prxmpt \cite{prxmpt}, and MDXPrompt \cite{mdxprompt}, focus on prompt authoring using JSX/TSX or MDX syntax, often integrating closely with specific runtime environments (e.g., Node.js) or prioritizing features like context length management via priority systems.
Specialized formats like ChatML \cite{chatml} define role-based structures for conversational history, while PromptML \cite{promptml} proposes a custom DSL focused on intention components but currently lacks integrated tooling.
SAMMO \cite{schnabel2024symbolicpromptprogramsearch} takes a distinct approach, using a Python API for programmatic manipulation and structure-aware optimization rather than direct markup authoring.

In contrast to these approaches, POML aims for a comprehensive, standalone solution distinguished by several key characteristics evident in the comparison.
It provides a language-agnostic runtime, enhancing portability compared to tools tied to Node.js or Python.
POML incorporates versatile components covering basic structure, intention expression, and data integration, offering arguably the most extensive built-in support among the listed tools for diverse data types like tables, documents, and webpages.
POML also integrates a unique CSS-like styling system for decoupling presentation from content and provides a full VS Code IDE extension, seeking to unify structure, data handling, styling, and development tooling within a single cohesive framework more completely than other approaches listed in \autoref{tab:prompt_tool_comparison}.

\section{Templating Engine Design Details}
\label{sec:poml_templating_detail}

To support the creation of dynamic and data-driven prompts, POML integrates a built-in templating engine.
This capability is inspired by widely adopted templating systems in web development frameworks \cite{django, jinja, twig, handlebars, jsx, angular, vuejs} and common practices in existing prompting frameworks \cite{langchain, microsoftguidance}.
Providing an \emph{integrated} engine enhances POML's utility, particularly for scenarios where prompts need to be generated dynamically based on runtime data without external scripting languages (\textbf{DG2}), contributing to efficient development workflows (\textbf{DG4}).

The engine allows for the insertion of variable values directly into the prompt text using a familiar double curly brace syntax: \texttt{{\{\{}variable{\}\}}}.
Its syntax for variables and control structures draws inspiration specifically from popular engines like Jinja and Handlebars \cite{jinja,handlebars}.
These variables can represent contextual placeholders that are populated at runtime from external data sources (e.g., documents, table records) or defined programmatically within the POML document itself using the \texttt{<let>} tag, enabling dynamic data binding and temporary variable assignment within the template logic.

Beyond simple variable substitution, the templating engine supports essential control structures for generating complex prompt logic dynamically.
Iteration over data collections (like lists or arrays) is handled using a \texttt{for} attribute, allowing repetitive structures to be generated concisely (e.g., \texttt{<list><item for="file in files">\{\{file.description\}\}</item></list>} would create a list item for each object in the \texttt{files} collection).
Conditional rendering of prompt sections is achieved through \texttt{if}/\texttt{else}-like constructs, enabling parts of the prompt to be included or excluded based on the evaluation of variable values or logical expressions (e.g., conditionally including a detailed explanation section only if a \texttt{verbose} flag is set to true).

The inclusion of this integrated templating engine offers several advantages for prompt engineering.
It significantly reduces redundancy by allowing developers to define reusable prompt structures that can be populated with different data inputs, rather than creating many similar static prompts.
Compared to constructing prompts through manual string manipulation or concatenation in a general-purpose programming language, which can be often error-prone and less readable \cite{carvalho2020text}, POML's templating approach makes the prompt structure more visually apparent and closer to a ``What You See Is What You Get'' (WYSIWYG) experience, simplifying the debugging process.
The engine's standalone capability provides dynamic prompting capabilities without mandating the use of an external programming language wrapper, differentiating POML from some other markup approaches that primarily serve as static data formats \cite{chatxml,promptml,mdxprompt}.

\section{Three-Pass Rendering Framework}
\label{sec:three_pass_rendering}

\begin{figure*}
  \centering
  \includegraphics[width=\linewidth]{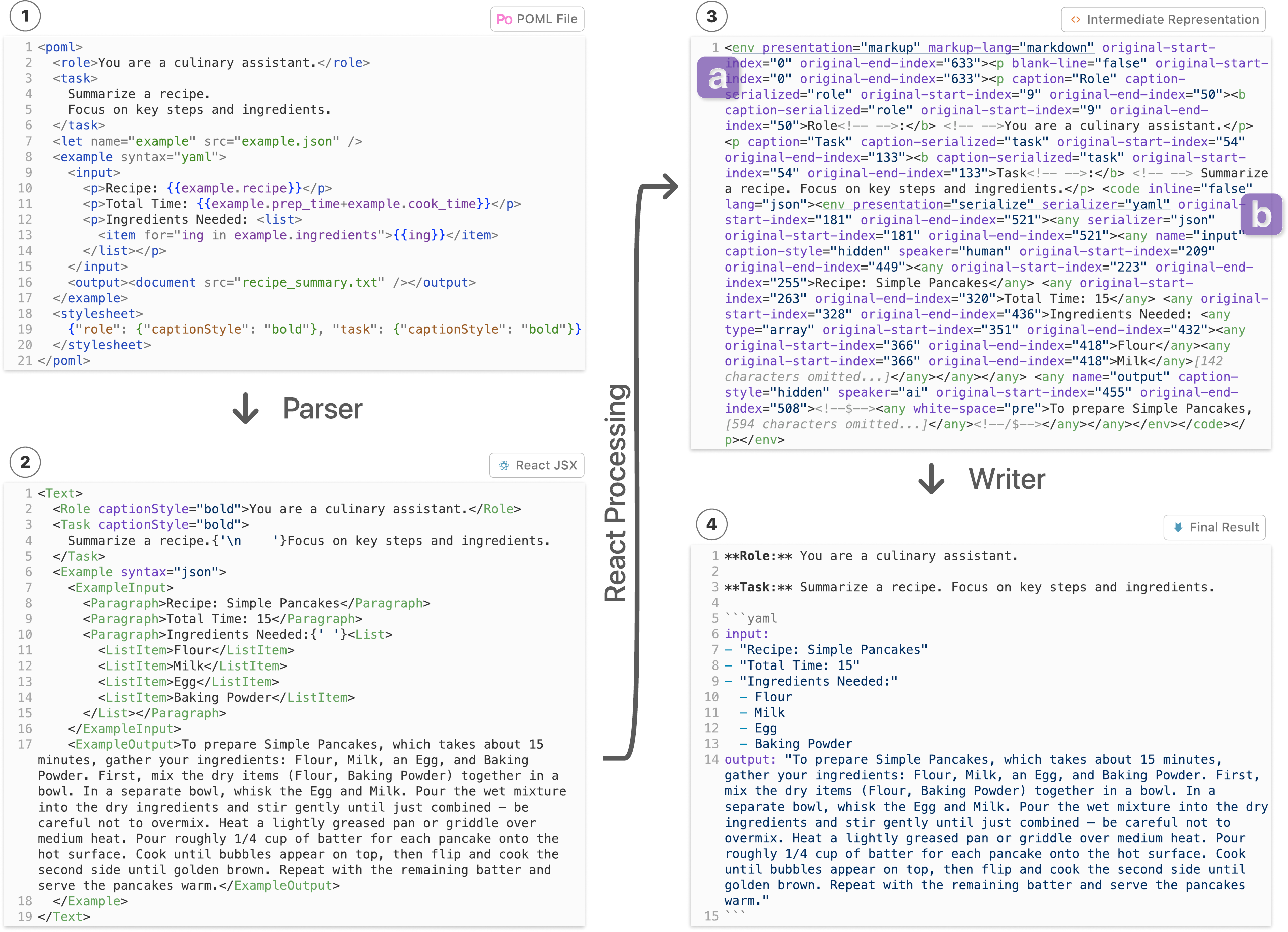}
  \caption{The POML three-pass rendering architecture: (1)->(2) The Parser transforms POML markup into React JSX components, (2)->(3) React processes these components into a detailed Intermediate Representation (IR) that captures content structure, styling, and metadata, and (3)->(4) specialized Writers convert the IR into various output formats such as JSON or Markdown.}
  \label{fig:implementation_flow}
\end{figure*}

The POML implementation employs a three-pass rendering framework.
This architecture processes the POML markup in three distinct stages: first, a ``Parser'' pass transforms the input into React JSX components; second, React processes these components to build a comprehensive Intermediate Representation (IR); and third, a ``Writer'' pass converts this IR into the desired final output format (e.g., Markdown, JSON).
The primary motivation for this three-pass approach is the clear separation of concerns.
The first pass (Parser) focuses exclusively on parsing the input markup, validating its structure, applying templating, stylesheets, and transforming it into React components.
The second pass (React processing) focuses on generating a detailed IR with complete metadata.
The third pass (Writer) is solely responsible for traversing the validated IR and serializing it into a specific target format.
This separation minimizes interdependencies, ensuring that changes related to styling or data handling do not inadvertently break the final rendering logic.

This three-tier separation fosters flexibility and extensibility.
Adding support for new output formats only requires implementing a new Writer module, without modifying the core parsing, templating, or styling logic.
Feature additions are simplified as each pass addresses a distinct, well-defined set of tasks.
From a performance perspective, the generated IR can be cached and reused to produce multiple output formats (e.g., generating both a plain text version and a chat-structured version simultaneously) without incurring the cost of re-parsing the original POML source, which is particularly beneficial for large prompts involving extensive data components.

The first pass, executed by the Parser module, begins by parsing the raw POML markup.
It identifies all known POML tags (e.g., \texttt{<task>}, \texttt{<document>}, \texttt{<table>}) and their attributes.
The parser validates the markup structure against the POML specification, flagging malformed tags or invalid attributes, often with a degree of error tolerance to handle minor issues without halting processing entirely (\S~\ref{sec:diagnostics}).
The template engine (\S~\ref{sec:poml_templating}) executes control flow attributes like \texttt{for}-loops and \texttt{if}-\texttt{else} conditions.
Variable substitutions using the \texttt{{\{\{}...{\}\}}} syntax are performed, resolving references to internal variables or data loaded from external sources (e.g., via \texttt{<let src="...">}).
Styling rules are applied from both the \texttt{<stylesheet>} tag and inline style attributes, resolving conflicts and computing final style properties for each element.
The Parser then transforms these validated elements into React JSX components, preparing them for the second pass.

In the second pass, React processes the JSX components generated by the Parser.
Using React's virtual DOM concept, this pass constructs a comprehensive Intermediate Representation (IR) - an in-memory tree of structured nodes corresponding to the POML elements, their resolved content, computed styles, and extensive metadata including original source positions and serialization hints.
This IR contains 21 distinct node types (Appendix~\ref{sec:ir_specification}) and captures all information needed for the final rendering stage, creating a clean, validated representation ready for the third pass.

The third pass is handled by a selected Writer module.
Its purpose is to convert the structured Intermediate Representation (IR) generated by React into a final, usable output format, such as Markdown, plain text, or a specific JSON structure required by an LLM API.

The Writer traverses the IR tree, visiting each node (e.g., an IR node representing a \texttt{<list>}, \texttt{<img>}, or \texttt{<document>}).
Based on the type of the IR node and the target output format, the Writer serializes the node's content appropriately.
For example, a \texttt{list} node might be rendered as a bulleted or numbered list in Markdown, while a \texttt{table} node could be formatted as CSV or a Markdown table depending on the Writer's configuration and the style attributes captured in the IR.
Furthermore, the Writer propagates and computes metadata associated with each node, such as the ``speaker'' attribute.
This allows the Writer to structure the output appropriately for different contexts, for instance, by splitting the final rendered text into distinct sections based on the speaker (e.g., ``human'', ``system'') if required by the target format (like chat completion APIs).

Different Writer modules can be implemented and selected to produce various outputs from the exact same IR.
This architecture makes it straightforward to extend POML to support new target formats (e.g., LaTeX, specific API JSON schemas) by simply creating a new Writer implementation without modifying the earlier passes.

Writers are designed with error tolerance in mind.
They can handle partially formed IR nodes (if errors occurred during earlier passes but were tolerated) or missing data references that could not be resolved.
If configuration parameters specific to the writer (e.g., a required CSV delimiter) are missing or invalid, the writer typically alerts the developer but attempts to continue converting the parts of the IR that are unaffected.

This three-pass structure consistently maintains the separation of concerns between prompt logic/content and final presentation/layout.
Updates to data sources (\texttt{src} attributes) or styling rules (\texttt{<stylesheet>}) captured during the earlier passes generally do not require any changes to the Writer logic.
This reinforces POML's commitment to a modular and robust approach to prompt engineering, aligning with design goals for Reusability (\textbf{DG1}) and Style Management (\textbf{DG3}).

\section{Intermediate Representation (IR) Specifications}
\label{sec:ir_specification}

\subsection{Attributes Applicable to All Tags}
The following attributes can be applied to any tag in the representation:
\begin{itemize}
  \item \texttt{speaker} (\textit{ai/human/system}): The speaker of the current content
  \item \texttt{original-start-index} (\textit{integer}): The start offset of the element corresponding to the current one in the original document
  \item \texttt{original-end-index} (\textit{integer}): The end offset of the element corresponding to the current one in the original document
\end{itemize}

\subsection{Tag Definitions}
\begin{description}[style=unboxed, leftmargin=!, itemsep=0.5em]
  \item[\texttt{any}] Represents a generic container for arbitrary data values. Useful for storing dynamic or unstructured content.
  \begin{itemize}
    \item \texttt{type} (\textit{string}): The data type of the value (`string', `integer', `float', `boolean', `array', `object', `buffer', `null', or `undefined').
    \item \texttt{name} (\textit{string}): An optional identifier for the data.
  \end{itemize}
  \item[\texttt{b}] Represents text that should be displayed in boldface. Useful for highlighting important words or phrases.
  \item[\texttt{code}] Represents a block or inline fragment of code. It can optionally include language and formatting attributes.
  \begin{itemize}
    \item \texttt{inline} (\textit{boolean}): Indicates whether the code is inline (true) or a block element (false).
    \item \texttt{lang} (\textit{string}): Specifies the programming language or syntax highlighting mode.
    \item \texttt{blank-line} (\textit{boolean}): Inserts a blank line before and after the code block if inline = false.
  \end{itemize}
  \item[\texttt{env}] Represents a formatting environment or container to specify how nested content should be output.
  \begin{itemize}
    \item \texttt{presentation} (\textit{string}): The output style or format mode (`markup', `serialize', `free', or `multimedia').
    \item \texttt{markup-lang} (\textit{string}): The specific markup language, required only if presentation = `markup'.
    \item \texttt{serializer} (\textit{string}): The name of the serializer, required only if presentation = `serialize'.
    \item \texttt{writer-options} (\textit{object}): Optional parameters passed to the writer constructor for customizing output.
  \end{itemize}
  \item[\texttt{h}] Represents a heading element.
  \begin{itemize}
    \item \texttt{level} (\textit{integer}): Indicates the heading level. Typically ranges from 1 (highest level) to 6 (lowest level).
  \end{itemize}
  \item[\texttt{i}] Represents text that should be displayed in italics. Useful for emphasizing words or phrases.
  \item[\texttt{img}] Represents an image element.
  \begin{itemize}
    \item \texttt{base64} (\textit{string}): The base64-encoded image data.
    \item \texttt{alt} (\textit{string}): Alternative text describing the image.
    \item \texttt{position} (\textit{string}): The placement of the image relative to text, such as `here', `top', or `bottom'.
    \item \texttt{type} (\textit{string}): The image MIME type (e.g., `image/jpeg', `image/png').
  \end{itemize}
  \item[\texttt{item}] Represents a single item within a list. Typically used as a child element of ``list''.
  \item[\texttt{list}] Represents an ordered or unordered list of items.
  \begin{itemize}
    \item \texttt{list-style} (\textit{string}): The style of the list bullets or enumeration (e.g., `star', `dash', `decimal').
  \end{itemize}
  \item[\texttt{nl}] Inserts newline characters.
  \begin{itemize}
    \item \texttt{count} (\textit{integer}): Specifies how many newline characters to insert.
  \end{itemize}
  \item[\texttt{obj}] Represents a data object, typically stored in JSON format.
  \begin{itemize}
    \item \texttt{data} (\textit{object}): A valid JSON object containing the structured data.
  \end{itemize}
  \item[\texttt{p}] Represents a paragraph of text. Useful for dividing content into readable blocks.
  \begin{itemize}
    \item \texttt{blank-line} (\textit{boolean}): Inserts a blank line before and after the paragraph when true.
  \end{itemize}
  \item[\texttt{s}] Represents text that should be displayed with a strikethrough style.
  \item[\texttt{span}] Represents an inline container for text without additional formatting. Useful for applying attributes without changing display structure.
  \item[\texttt{table}] Represents a table structure containing rows and cells.
  \item[\texttt{tbody}] Represents the body section of a table, containing the majority of data rows.
  \item[\texttt{tcell}] Represents a single cell within a table row.
  \item[\texttt{text}] Represents raw or unformatted text content.
  \item[\texttt{thead}] Represents the header section of a table, typically containing column headings.
  \item[\texttt{trow}] Represents a single row within a table, containing one or more cells.
  \item[\texttt{u}] Represents text that should be displayed with an underline.
\end{description}

\section{Details of TableQA Case Study}
\label{sec:tableqa_prompt}

\begin{figure}[t]
  \includegraphics[width=\linewidth]{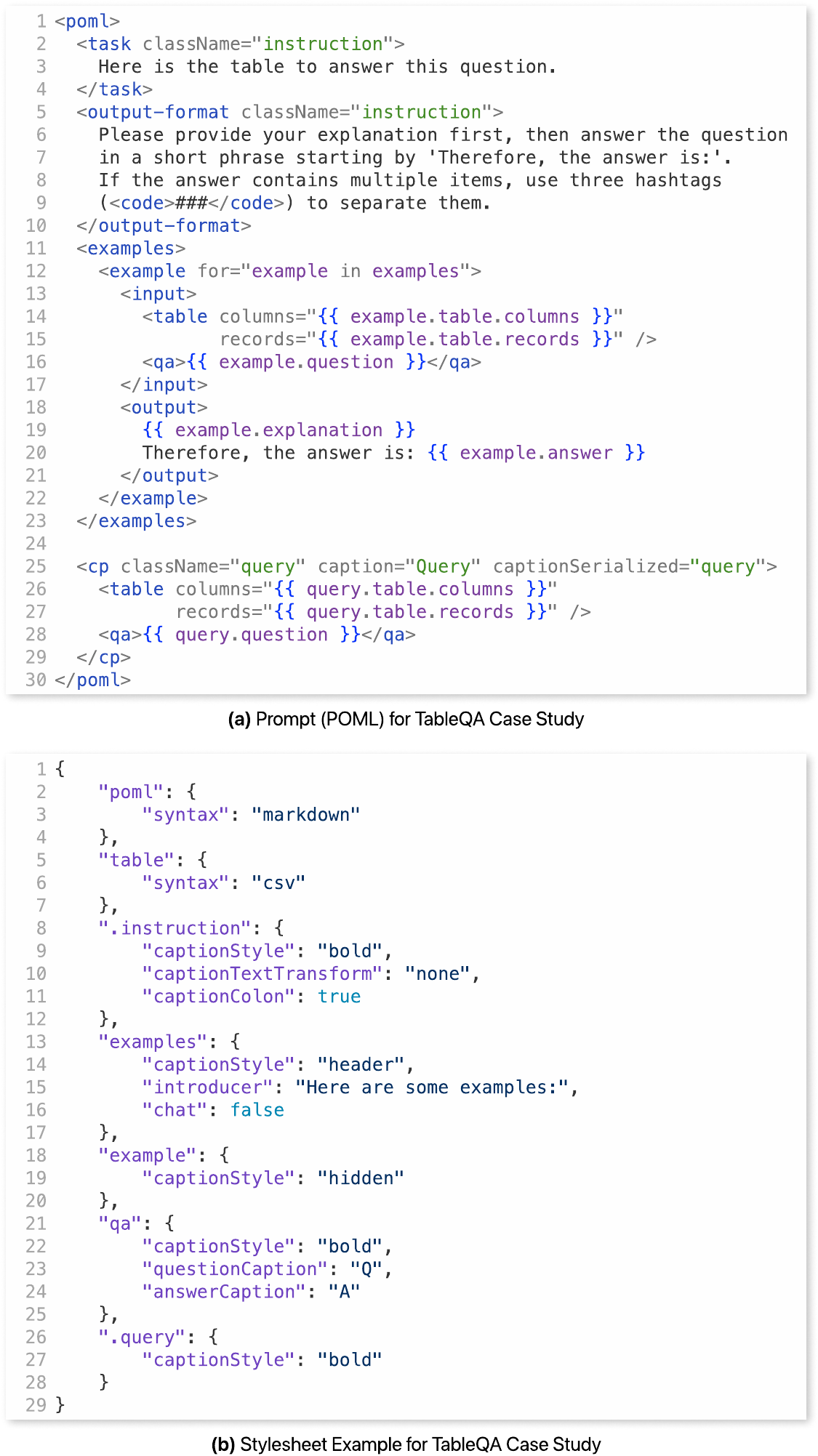}
  \caption{Example of POML usage for the TableQA case study (\S~\ref{sec:cs_tableqa}, Appendix~\ref{sec:tableqa_prompt}). (a) The POML prompt template defining the task instructions, required output format, few-shot examples, and the query structure containing the table and question. (b) The corresponding JSON stylesheet example specifying formatting and presentation rules for elements within the POML prompt, such as overall syntax, caption styles, and introducer text.}
  \label{fig:tableqa_prompt}
\end{figure}

A detailed example of the prompt structure used in our TableQA case study is provided below, as discussed in the main text. Figure~\ref{fig:tableqa_prompt} illustrates the components involved.

Figure~\ref{fig:tableqa_prompt}(a) presents the POML template crafted for this task. It defines the overall structure provided to the language model, encompassing the core task instruction, specific requirements for the output format (including explanation prefixes and separators), few-shot examples demonstrating the desired input-output behavior, and the placeholder for the actual query which includes the table data (columns and records) and the user question.

Figure~\ref{fig:tableqa_prompt}(b) shows an example of the corresponding JSON stylesheet. This stylesheet controls the rendering and formatting aspects of the POML prompt, specifying details such as the syntax for different elements (e.g., markdown for POML, csv for tables), styling for instructional text and captions (e.g., bolding, headers, specific question/answer prefixes like ``Q''-``A''), and the presentation of examples. This separation of content (POML) and presentation (stylesheet) allows for flexible prompt customization and management.

\begin{table*}
  \caption{Impact analysis of specific prompt styling choices on TableQA accuracy across different LLMs (\S~\ref{sec:cs_tableqa}, Appendix~\ref{sec:tableqa_prompt}). Each cell indicates whether a feature (row) significantly improved (\mycheckmark), had no significant effect (\mydash), or significantly harmed (\mycross) performance for a given LLM (column), based on statistical significance tests (Mann-Whitney U test comparing styles with vs. without the feature; significance levels indicated). The bottom rows show examples of interaction effects for specific feature combinations.}
  \label{tab:feature_pref}
  \resizebox{\textwidth}{!}{


\begin{tabular}{l|cccccccc}
\hline
\textbf{Style Preference} & \textbf{Claude} & \textbf{DeepSeek} & \textbf{Gemini} & \textbf{3.5-Turbo} & \textbf{4o-Mini} & \textbf{LLaMA-3} & \textbf{Mistral} & \textbf{Phi-3} \\ \hline
Example Body = Text & \pref{$p <$ 0.01} & \pref{$p <$ 0.05} & \nopref & \nopref & \pref{$p <$ 0.01} & \notpref{$p <$ 0.01} & \notpref{$p <$ 0.01} & \pref{$p <$ 0.01} \\
Example Body = Introducer & \pref{$p <$ 0.01} & \nopref & \pref{$p <$ 0.01} & \notpref{$p <$ 0.01} & \pref{$p <$ 0.01} & \nopref & \notpref{$p <$ 0.01} & \pref{$p <$ 0.01} \\
Overall Syntax = HTML & \nopref & \nopref & \pref{$p <$ 0.05} & \notpref{$p <$ 0.01} & \pref{$p <$ 0.01} & \nopref & \notpref{$p <$ 0.05} & \pref{$p <$ 0.01} \\
Overall Syntax = XML & \pref{$p <$ 0.05} & \nopref & \nopref & \pref{$p <$ 0.05} & \nopref & \notpref{$p <$ 0.05} & \notpref{$p <$ 0.01} & \nopref \\
Table Syntax = CSV & \pref{$p <$ 0.05} & \notpref{$p <$ 0.01} & \notpref{$p <$ 0.01} & \nopref & \nopref & \nopref & \nopref & \pref{$p <$ 0.01} \\
Overall Syntax = Markdown & \nopref & \nopref & \nopref & \pref{$p <$ 0.01} & \notpref{$p <$ 0.01} & \nopref & \pref{$p <$ 0.01} & \notpref{$p <$ 0.01} \\
Example Body = Chat & \notpref{$p <$ 0.01} & \nopref & \nopref & \nopref & \notpref{$p <$ 0.01} & \pref{$p <$ 0.01} & \pref{$p <$ 0.01} & \notpref{$p <$ 0.01} \\
Example Body = Chat w. Introducer & \notpref{$p <$ 0.01} & \notpref{$p <$ 0.05} & \notpref{$p <$ 0.05} & \nopref & \notpref{$p <$ 0.01} & \pref{$p <$ 0.05} & \pref{$p <$ 0.01} & \notpref{$p <$ 0.01} \\
QA Caption = Hidden & \notpref{$p <$ 0.01} & \nopref & \notpref{$p <$ 0.01} & \pref{$p <$ 0.05} & \notpref{$p <$ 0.01} & \nopref & \nopref & \notpref{$p <$ 0.01} \\
Overall Syntax = Markdown, Example Body = Introducer & \pref{$p <$ 0.01} & \pref{$p <$ 0.05} & \pref{$p <$ 0.05} & \nopref & \pref{$p <$ 0.05} & \nopref & \notpref{$p <$ 0.01} & \pref{$p <$ 0.01} \\
Instruction Header = Header, Example Body = Introducer & \pref{$p <$ 0.05} & \nopref & \pref{$p <$ 0.05} & \notpref{$p <$ 0.01} & \pref{$p <$ 0.05} & \nopref & \nopref & \pref{$p <$ 0.05} \\
Example Body = Introducer, QA Caption = Bold-Upper & \pref{$p <$ 0.05} & \nopref & \pref{$p <$ 0.05} & \nopref & \pref{$p <$ 0.05} & \nopref & \notpref{$p <$ 0.05} & \pref{$p <$ 0.01} \\
Example Body = Introducer, QA Body = Question-Explanation & \pref{$p <$ 0.01} & \nopref & \pref{$p <$ 0.05} & \notpref{$p <$ 0.05} & \pref{$p <$ 0.01} & \nopref & \nopref & \pref{$p <$ 0.01} \\
Example Body = Text, QA Body = Question-Explanation & \pref{$p <$ 0.01} & \pref{$p <$ 0.05} & \nopref & \pref{$p <$ 0.05} & \pref{$p <$ 0.01} & \notpref{$p <$ 0.01} & \notpref{$p <$ 0.01} & \pref{$p <$ 0.05} \\
Overall Syntax = Markdown, Example Body = Text & \pref{$p <$ 0.01} & \pref{$p <$ 0.05} & \nopref & \pref{$p <$ 0.01} & \pref{$p <$ 0.01} & \notpref{$p <$ 0.01} & \notpref{$p <$ 0.01} & \pref{$p <$ 0.01} \\
\hline
\end{tabular}
}
\end{table*}

\begin{figure}[t]
  \centering
  \includegraphics[width=\linewidth]{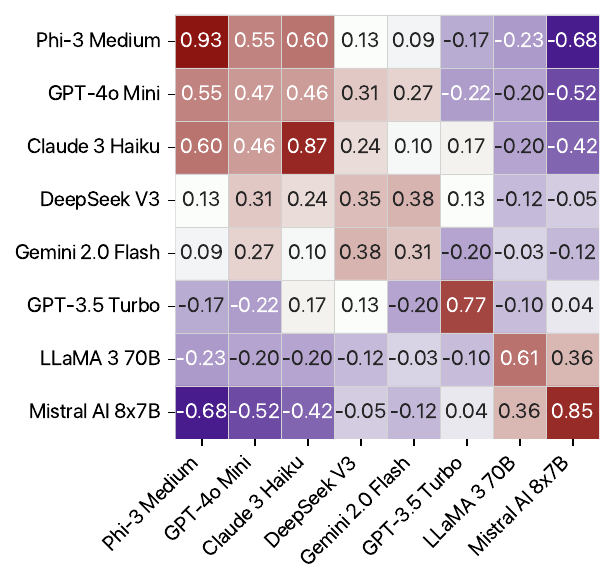}
  \caption{Correlation matrix showing relationships between LLMs based on their performance rankings across 100 prompt styles in the TableQA task (\S~\ref{sec:cs_tableqa}, Appendix~\ref{sec:tableqa_prompt}). Cells show Spearman correlation coefficients between model style rankings (Red: positive/similar preferences; Blue: negative/dissimilar preferences). The diagonal shows the self-correlation score for each model (see \autoref{tab:model_perf_best} for definition and values), indicating style ranking stability. The self-correlation score indicates the stability of the style performance ranking. It is computed by randomly splitting the 283 samples into two equal halves, calculating the accuracy of each of the 100 styles on both halves, and finding the Spearman correlation between the two resulting style rankings. This process is repeated 1000 times, and the mean correlation is reported; higher values indicate more stable style rankings across different data subsets.}
  \label{fig:correlation_heatmap}
\end{figure}

Analysis of style preferences across models, visualized in the correlation heatmap (\autoref{fig:correlation_heatmap}), revealed distinct model groupings.
For instance, Phi-3 Medium, GPT-4o Mini, and Claude 3 Haiku formed a cluster exhibiting similar positive correlations in their style rankings, suggesting shared preferences.
Conversely, Mistral AI 8x7B and LLaMA 3 70B formed another group, often showing negative correlations with the first cluster, indicating fundamentally different sensitivities to styling choices.
Deeper analysis of individual feature impacts (\autoref{tab:feature_pref}) highlighted specific sensitivities with statistical significance.
Using plain text for example bodies significantly benefited models like Claude 3 Haiku and Phi-3 Medium but harmed LLaMA 3 70B and Mistral AI 8x7B ($p < 0.01$).
Adopting a chat-like format for examples helped LLaMA 3 70B and Mistral AI 8x7B but was detrimental to several others ($p < 0.01$).
Hiding structural elements like QA captions negatively impacted most models but surprisingly benefited GPT-3.5 Turbo ($p < 0.05$).
Complex interaction effects were also observed, where the combined impact of certain style features differed from the sum of their individual effects, indicating non-linear relationships in how styling elements influence performance.

\section{Semi-structured Verbal Interview Questions}
\label{sec:interview_questions}

The following questions were used to guide the semi-structured interviews with users.
The questions were designed to elicit detailed feedback on the user's experience with POML, its features, and their suggestions for improvement.

\begin{enumerate}
\item What tasks do you find most challenging and why?
\item Do you have prior experience with using prompts? In what scenarios do you typically employ prompts?
\item In your opinion, for which scenarios is POML most useful?
\item In your opinion, for which scenarios is POML not useful?
\item Which parts or features of POML do you find most valuable, and why?
\item Which parts or features of POML do you find most difficult to use, and why?
\item Do you consider the POML syntax easy to learn and worthwhile to master?
\item How do you foresee integrating POML into your existing workflow or pipeline?
\item Do you like the VSCode integration of POML (preview, testing, auto-completion)? Why or why not?
\item What improvements or additional features do you suggest for POML or its VSCode integration?
\item Did you encounter any bugs or errors, and how did these affect your overall experience of POML?
\end{enumerate}


\end{document}